\documentclass[prb,reprint,a4paper,showpacs,citeautoscript,floatfix]{revtex4-1}

\pdfoutput=1
\usepackage[charter]{mathdesign}
\usepackage{amsmath}

\usepackage[pdftex]{graphicx}
\usepackage{microtype}
\usepackage{bm}\let\vec\bm

\usepackage[svgnames]{xcolor}
\usepackage[
	colorlinks=True,linkcolor=DarkRed,citecolor=ForestGreen,urlcolor=MediumBlue,
	pdfstartview=FitH,bookmarks=False,pdfpagemode=UseNone
]{hyperref}

\def\frequency{\omega_{\mathrm{D}}}
\def\cutoff{\hbar\frequency}
\def\coupling{\bar{\lambda}}

\begin{document}

\title{BCS superconductivity near the band edge: Exact results for one and several bands}

\author{D. Valentinis}
\author{D. van der Marel}
\author{C. Berthod}
\email[To whom correspondence should be addressed. E-mail: ]{christophe.berthod@unige.ch}
\affiliation{Department of Quantum Matter Physics (DQMP), University of Geneva, 24 quai Ernest-Ansermet, 1211 Geneva 4, Switzerland}

\date{January 19, 2016}

\begin{abstract}

We revisit the problem of a BCS superconductor in the regime where the Fermi energy is smaller than the Debye energy. This regime is relevant for low-density superconductors such as SrTiO$_3$ that are not in the BEC limit, as well as in the problem of ``shape resonances'' associated with the confinement of a three-dimensional superconductor. While the problem is not new, exact results were lacking in the low-density limit. In two dimensions, we find that the initial rise of the pairing temperature $T_c$ at low density $n$ is nonanalytic and faster than any power of $n$. In three dimensions, we also find that $T_c$ is nonanalytic, but starts with \emph{zero} slope at weak coupling and infinite slope at strong coupling. Self-consistent treatment of the chemical potential and energy dependence of the density of states are crucial ingredients to obtain these results. We also present exact results for multiband systems and confirm our analytical expressions by numerical simulations.

\end{abstract}

\maketitle

\section{Introduction}

The Bardeen--Cooper--Schrieffer (BCS) theory \cite{Bardeen-1957} remains the only strong microscopic foundation to support our understanding of the fascinating phenomenon of superconductivity. Among many other insights, the theory provides a simple expression for the critical temperature $T_c$, which continues to inspire the search for materials with improved performances. In particular, it is expected that superconductivity is favored by a low dimensionality due to enhanced density of states (DOS) at the Fermi level \cite{Eagles-1967}. For three-dimensional (3D) materials, an early proposal to use quantum confinement in a thin film \cite{Thompson-1963} has received sustained attention until recently \cite{[{See, e.g., }][{, and references therein.}] Romero-Bermudez-2014}. The confinement-induced two-dimensional (2D) subbands produce discontinuities in the DOS and abrupt changes of $T_c$ as a function of film thickness have been routinely predicted.

\begin{figure}[b]
\includegraphics[width=0.6\columnwidth]{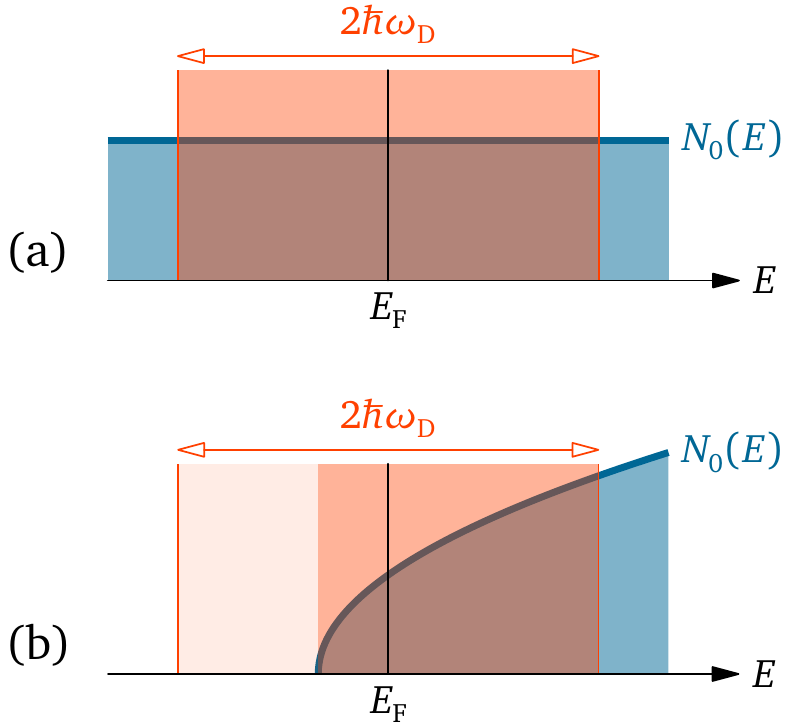}
\caption{\label{fig:band-edge}
Schematic representation of (a) the high-density regime and (b) the low-density regime for superconducting pairing. In the former, $E_{\mathrm{F}}\gg\cutoff$ and the density of states $N_0(E)$ can be taken constant. In the latter, $E_{\mathrm{F}}\lesssim\cutoff$, the interaction is cut by the band edge, and the details of $N_0(E)$ matter.
}
\end{figure}

The purpose of this study is to explore some consequences of an aspect of the problem, considered by Eagles half a century ago \cite{Eagles-1967, Eagles-1969a, *Eagles-1969b}, but often overlooked in recent calculations based on the BCS gap equation. As the Fermi energy crosses the edge of a band, there is a regime where the dynamical cutoff of the pairing interaction is controlled by the band edge (Fig.~\ref{fig:band-edge}). This regime is realized in low-density electron gases, when the Fermi energy is smaller than the dynamical range of the interaction. In doped SrTiO$_3$, for instance, the carrier concentration is typically $10^{19}~\mathrm{cm}^{-3}$ and the carrier mass is in the range 2--4 electronic masses \cite{Lin-2014}, corresponding to a Fermi temperature of $50$--$100$~K, while the Debye temperature is $513$~K \cite{Ahrens-2007}. In this situation, the common approximation of taking a constant DOS over the full dynamical range fails to give a good estimate for $T_c$. The near-band edge regime is also relevant in the quasi-2D problem of shape resonances, since each resonance is due to the Fermi energy crossing a subband edge \cite{Perali-1996, Innocenti-2010, *Innocenti-2011, Guidini-2014}. The pairing in that subband, as well as the inter-subband pairing involving that subband, are dominated by the band edge. A synthesis of these two cases is realized in the quasi-2D \emph{and} low-density electron gas at the LaAlO$_3$/SrTiO$_3$ interface \cite{Zubko-2011, Pentcheva-2010, Cancellieri-2014}. In the present paper we focus on the band-edge effect on $T_c$ in the bulk, emphasizing the generic behaviors in the simple case of an electron gas with parabolic dispersion and a local attraction. We recover the expressions of Eagles \cite{Eagles-1967} in the weak-coupling limit. In the low-density regime, we provide exact relations as a function of the density, which are valid at arbitrary coupling. We also give exact numerical results in 2D and 3D, for one-band and multiband systems. The implications for the problem of quasi-2D shape resonances and the case of LaAlO$_3$/SrTiO$_3$, will be reported in separate publications.

Our starting point is the mean-field theory for a momentum-independent pairing interaction acting in a limited energy range around the Fermi surface. This theory yields a pairing temperature which is in general higher than the temperature of superconducting coherence, especially when the dimensionality and/or the density is low and superconducting fluctuations become important \cite{*[{}] [{ [Sov. Phys. JETP \textbf{32}, 493 (1971)].}] Berezinskii-1970, Kosterlitz-1973, Larkin-2005}. We ignore these fluctuations and focus on the mean-field equations, refraining from making any approximation when solving them for $T_c$. This approach is similar to previous mean-field studies of the BCS-BEC crossover where the renormalized chemical potential is solved self-consistently together with the $T_c$ equation or the gap equation at zero temperature \cite{Leggett-1980, [{For a recent review, see }] Randeria-2014}.

An exact solution of the gap equation requires one to take into account the energy dependence of the DOS, most importantly the cutoff at the band bottom, and the temperature dependence of the chemical potential $\mu$, which is crucial at low-density $n$. Because $n$, $T_c$, and $\mu$ all approach zero simultaneously, it is essential to use the exact relation $\mu(n,T_c)$ in order to capture the correct behavior of $T_c$ for $n\to0$. Furthermore, one should not assume weak coupling and/or assume that $T_c$ is small with respect to the Fermi energy and the cutoff for pairing. As a matter of fact, analytical results in this problem are rare. In Ref.~\onlinecite{Hainzl-2008}, rigorous bounds for $T_c$ were obtained for a general interaction. These results are limited to weak coupling and to a positive chemical potential. We will see that the chemical potential at $T_c$ is negative in the low-density limit in 2D for any coupling and in 3D for couplings larger than a critical value. Exact results have also been reported for the zero-temperature gap in 2D \cite{vanderMarel-1990}. However, since the universal BCS gap to $T_c$ ratio is \emph{not} obeyed in the low-density limit, these results cannot be used to deduce $T_c$.

This paper is the first in a series and it provides the mathematical foundations for subsequent studies dedicated to shape resonances in thin films and to the LaAlO$_3$/SrTiO$_3$ interface. It is organized as follows. In Sec.~\ref{sec:basics} we recall the basic coupled equations giving $n$ and $T_c$ and we write them in a dimensionless form, for one and several parabolic bands. In Sec.~\ref{sec:1band} we present our analytical and numerical results for one band in 2D and 3D, and in Sec.~\ref{sec:2bands} we briefly discuss multiband effects.

\section{\boldmath BCS $T_c$ equation for multiband systems}
\label{sec:basics}

\subsection{Dimensionless equations for the pairing temperature}

We consider a multiband metal with a local BCS pairing interaction $-V_{\alpha\beta}$ acting between electrons of opposite momenta and spins in bands $\alpha$ and $\beta$.\footnote{We pull out a minus sign from the interaction for convenience. $V_{\alpha\beta}$ can be positive (attractive interaction) or negative (repulsive interaction).} We assume that Cooper pairing occurs only for two electrons in the same band, leading below the pairing temperature $T_c$ to an order parameter $\Delta_{\alpha}$ in each band. This includes the possibility of a ``proximity'' induced gap $\Delta_{\beta}$ in a band that otherwise feels no pairing potential ($V_{\beta\beta}=0$), via the nonzero interband interactions $V_{\alpha\beta}$. The mean-field gap equation for $\Delta_{\alpha}$ is
	\begin{subequations}
	\begin{equation}
		\label{eq:gap-equationa}
		\Delta_{\alpha}=\sum_{\beta}V_{\alpha\beta}\Delta_{\beta}
		\int_{-\cutoff}^{\cutoff}\hspace{-2mm}d\xi\,N_{0\beta}(\mu+\xi)\,
		\frac{\tanh\left(\frac{\sqrt{\xi^2+\Delta_{\beta}^2}}{2k_{\text{B}}T}\right)}
		{2\sqrt{\xi^2+\Delta_{\beta}^2}}.
	\end{equation}
The pairing interaction acts in a range $\pm\cutoff$ around the chemical potential $\mu$. Although the notation $\cutoff$ is used here, we envision the problem in its generality and our results do not require phonon-mediated pairing, but apply to any local interaction with a dynamical cutoff. In a lattice version, for instance, the cutoff could be the bandwidth. $N_{0\beta}(E)$ is the DOS per spin and per unit volume for the band $\beta$. It is defined on an absolute energy scale, such that $N_{0\beta}(\mu)$ is the DOS at the chemical potential $\mu$, which is common to all bands. The chemical potential must be adjusted to fix the density according to 
	\begin{equation}
		n=2\int_{-\infty}^{\infty}dE\,f(E)\,N(E).
	\end{equation}
	\end{subequations}
Here, $f(E)=[e^{(E-\mu)/k_{\text{B}}T}+1]^{-1}$ is the Fermi distribution function and $N(E)$ is the total BCS density of states (per spin) resulting from the opening of the superconducting gaps at the chemical potential in each band.

For the calculation of $T_c$, it is sufficient to consider the two equations in the limit of vanishing order parameters. For $T=T_c$ we have
	\begin{subequations}\label{eq:general}
	\begin{align}
		\Delta_{\alpha}&=\sum_{\beta}V_{\alpha\beta}\Delta_{\beta}
		\int_{-\cutoff}^{\cutoff}dE\,N_{0\beta}(\mu+E)\,
		\frac{\tanh\left(\frac{E}{2k_{\text{B}}T_c}\right)}{2E}\\
		n&=2\int_{-\infty}^{\infty}dE\,f(E)\,\sum_{\beta}N_{0\beta}(E).
	\end{align}
	\end{subequations}
We now insert explicit formulas for the energy-dependent densities of states and the density and we rewrite the equations (\ref{eq:general}) in a dimensionless form, which is more convenient for analytical and numerical treatments. The densities of states for a parabolic band in dimensions $d=2$ and $d=3$ are given by
	\begin{equation}\label{eq:DOS_d}
		N_{0\beta}(E)=(d-1)\pi\left(\frac{m_{\beta}}{2\pi^2\hbar^2}\right)^{\frac{d}{2}}
		\theta(E-E_{0\beta})\left(E-E_{0\beta}\right)^{\frac{d}{2}-1},
	\end{equation}
where $m_{\beta}$ is the band mass, $E_{0\beta}$ is the energy of the band minimum, and $\theta$ is the Heaviside function. This definition ensures that $N_{0\beta}(\mu)$ is the DOS evaluated at the chemical potential $\mu$ common to all bands, consistently with Eq.~(\ref{eq:gap-equationa}). The relation between density, chemical potential, and temperature for a parabolic band in arbitrary dimension $d$ is
	\begin{equation}\label{eq:density}
		n=-2\left(\frac{mk_{\text{B}}T}{2\pi\hbar^2}\right)^{\frac{d}{2}}
		\text{Li}_{\frac{d}{2}}\left(-e^{\frac{\mu-E_0}{k_{\text{B}}T}}\right),
	\end{equation}
where $\text{Li}_{p}(x)$ is the polylogarithm given by the series expansion $\text{Li}_{p}(x)=\sum_{q=1}^{\infty}x^q/q^p$. This function has the sign of its argument and reduces to a usual logarithm in two dimensions ($p=1$): $\text{Li}_1(x)=-\ln(1-x)$. We provide a brief derivation of Eq.~(\ref{eq:density}) in Appendix~\ref{app:density} for the interested reader.

We measure all energies in units of $\cutoff$, express the density in units of $2[m\frequency/(2\pi\hbar)]^{d/2}$ where $m$ is a reference mass, and we distinguish the dimensionless variables with tildes, e.g.,
	\[
		\tilde{T}_c=\frac{k_{\mathrm{B}}T_c}{\cutoff},\quad
		\tilde{\mu}=\frac{\mu}{\cutoff},\quad
		\tilde{n}=\frac{n}{2[m\frequency/(2\pi\hbar)]^{d/2}},
		\quad\mathrm{etc.}
	\]
The coupled equations (\ref{eq:general}) for $T_c$ become
	\begin{subequations}\label{eq:dimensionless}
	\begin{align}
		\label{eq:dimensionlessa}
		\Delta_{\alpha}&=\sum_{\beta}\Delta_{\beta}\coupling_{\alpha\beta}\,
		\psi_d\left(1+\tilde{\mu}-\tilde{E}_{0\beta},\tilde{T}_c\right)\\
		\label{eq:dimensionlessb}
		\tilde{n}&=-\tilde{T}_c^{\frac{d}{2}}\sum_{\beta}
		\left(\frac{m_{\beta}}{m}\right)^{\frac{d}{2}}\text{Li}_{\frac{d}{2}}\left(
		-e^{\frac{\tilde{\mu}-\tilde{E}_{0\beta}}{\tilde{T}_c}}\right).
	\end{align}
	\end{subequations}
We have introduced the dimensionless function,
	\begin{equation}\label{eq:psi}
		\psi_d(a,b)=\theta(a)\int_{1-\min(a,2)}^1\hspace{-2em}dx\,(x+a-1)^{\frac{d}{2}-1}
		\frac{\tanh\left(\frac{x}{2b}\right)}{2x},
	\end{equation}
as well as the coupling constants,
	\begin{equation}\label{eq:lambda}
		\coupling_{\alpha\beta}=V_{\alpha\beta}(d-1)\pi
		\left(\frac{m_{\beta}}{2\pi^2\hbar^2}\right)^{\frac{d}{2}}
		\left(\cutoff\right)^{\frac{d}{2}-1}.
	\end{equation}
We use a bar to recall that these coupling constants are not evaluated at the Fermi energy like in the common practice, but at an energy $\cutoff$ above the bottom of each band: $\coupling_{\alpha\beta}=V_{\alpha\beta}N_{0\beta}(E_{0\beta}+\cutoff)$. This choice is natural and leads to the simplest equations. The usual definition $\lambda=VN_{0}(\mu)$ poses problems when $\mu$ lies below the band bottom and more generally because $\mu$ is a function of interaction strength and temperature.

In an $N$-band system, the relations (\ref{eq:dimensionless}) provide $N+1$ equations for the $N+1$ unknowns, which are $\tilde{T}_c$, $\tilde{\mu}$, and the $N-1$ ratios $\Delta_{\beta}/\Delta_1$. We can assume that $\Delta_1\neq0$ without loss of generality, because there is at least one nonzero gap parameter at $T_c$ and we are free to number the bands such that $\Delta_1$ is this one. We now eliminate the $N-1$ gap ratios and reduce the problem to a pair of equations for $\tilde{T}_c$ and $\tilde{\mu}$. With the new definitions $r_{\beta}=\Delta_{\beta}/\Delta_1$ and
	\begin{equation}
		\Lambda_{\alpha\beta}(\tilde{\mu},\tilde{T}_c)=\coupling_{\alpha\beta}
		\psi_d(1+\tilde{\mu}-\tilde{E}_{0\beta},\tilde{T}_c),
	\end{equation}
the set of $N$ equations (\ref{eq:dimensionlessa}) becomes the eigenvalue problem $\Lambda\vec{r}=\vec{r}$ with $\vec{r}=(1,r_2,\ldots,r_N)$. This means that, when evaluated at a value of $\tilde{T}_c$ solving Eq.~(\ref{eq:dimensionlessa}), the matrix $\Lambda$ has at least one unit eigenvalue. In other words, $\tilde{T}_c$ corresponds to the largest temperature that satisfies the characteristic equation $\det(\openone-\Lambda)=0$. The two coupled dimensionless equations giving $n$ and $T_c$ for $N$ bands are therefore
	\begin{subequations}\label{eq:tc-Nbands}
	\begin{align}
		\label{eq:tc-Nbandsa}
		0&=\det[\openone-\Lambda(\tilde{\mu},\tilde{T}_c)]\\
		\label{eq:tc-Nbandsb}
		\tilde{n}&=-\tilde{T}_c^{\frac{d}{2}}\sum_{\alpha=1}^N
		\left(\frac{m_{\alpha}}{m}\right)^{\frac{d}{2}}\,\text{Li}_{\frac{d}{2}}\left(
		-e^{\frac{\tilde{\mu}-\tilde{E}_{0\alpha}}{\tilde{T}_c}}\right).
	\end{align}
	\end{subequations}
The existence of a nontrivial solution to Eq.~(\ref{eq:dimensionlessa}) clearly implies Eq.~(\ref{eq:tc-Nbandsa}). The converse is also true: The vanishing of the determinant in Eq.~(\ref{eq:tc-Nbandsa}) is sufficient to enforce that the matrix $\Lambda$ has one unit eigenvalue, which provides a solution to Eq.~(\ref{eq:dimensionlessa}). The equations (\ref{eq:tc-Nbands}) have the same structure in 2D and 3D, the quantitative differences stemming mostly from different functions $\psi_d(a,b)$. In the next paragraph we discuss the properties of these functions, which we shall use in the following sections to derive analytical results.

\subsection{\boldmath Properties of the functions $\psi_d(a,b)$}

\begin{figure}[b]
\includegraphics[width=0.85\columnwidth]{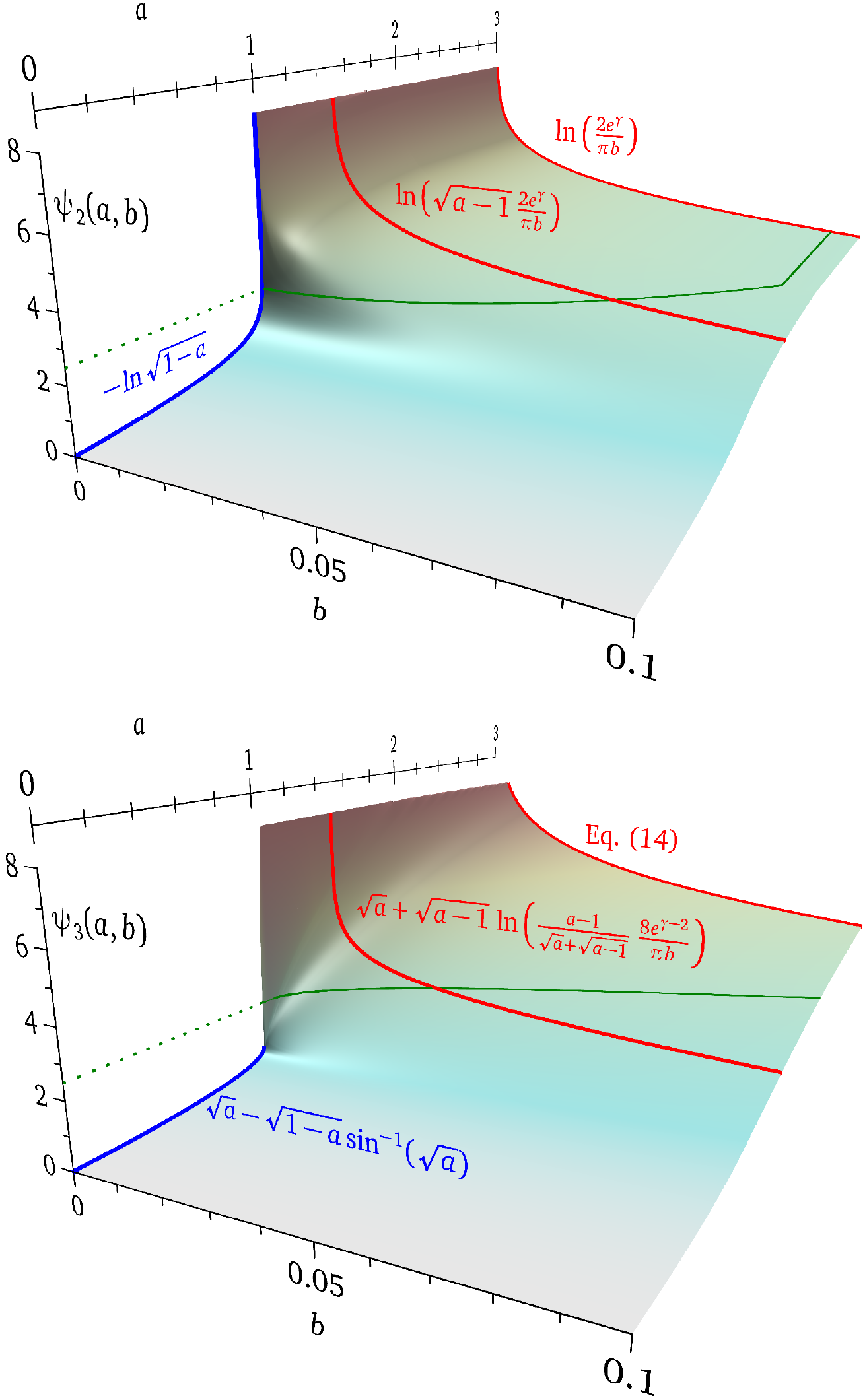}
\caption{\label{fig:psi}
Representation of the functions $\psi_d(a,b)$ defined in Eq.~(\ref{eq:psi}) for dimensions $d=2$ (top) and $d=3$ (bottom). Physically, the $a$ axis corresponds to varying $\mu$ around the band bottom ($a=1$) and the $b$ axis is proportional to $T_c$. The blue lines show the behavior for $b=0$ and $a<1$. The two red lines in each graph show the asymptotic $b$ dependencies for $1<a<2$ and $a>2$, respectively. The green lines show cuts at the value $\psi_d(a,b)=2.5$, which correspond to the path followed in the $(a,b)$ plane by the solution of the BCS equations (\ref{eq:tc-Nbands}) for one band and for $\coupling=0.4$.
}
\end{figure}

The functions $\psi_d(a,b)$ are displayed \footnote{The standard numerical integration packages fail to produce an accurate result for Eq.~(\ref{eq:psi}), or take a prohibitively long time to converge, especially in the limit $b\to0$ of interest to us. For a fast and accurate numerical evaluation of Eq.~(\ref{eq:psi}) we have used the representation of the Fermi function proposed in Ref.~\onlinecite{Osaki-2007}. With 14 Osaki poles and residues, we build a representation of the function $\tanh(x/2)$ which has an accuracy of $\sim10^{-16}$ for all $x$, setting the function to $\pm 1$ for $|x|>16\ln(10)$. With this representation, Eq.~(\ref{eq:psi}) can be evaluated analytically. This provides a very fast implementation of the functions $\psi_d(a,b)$ with double precision accuracy.}\phantom{\onlinecite{Osaki-2007}\hspace{-1.2em}} in Fig.~\ref{fig:psi}. The strongest structure develops around $a=1$, which corresponds physically to having the chemical potential at the bottom of one band. We are mostly interested in the behavior for $b\ll1$, which is explored in the regime $k_{\mathrm{B}}T_c\ll\cutoff$ and particularly in the limit $b\to0$, which is relevant when the density approaches zero. If $a<1$, $\psi_d(a,b)$ is finite for $b=0$. The limiting value is given by
	\begin{multline}\label{eq:psialess1}
		\psi_d(0<a<1,b\to 0)=\int_{\frac{1-a}{b}}^{\frac{1}{b}}dx\,
		\frac{(bx+a-1)^{\frac{d}{2}-1}}{2x}\\
		=\begin{cases}-\ln\sqrt{1-a}&\quad(d=2)\\[1em]\sqrt{a}-\sqrt{1-a}\sin^{-1}(\sqrt{a})
		&\quad(d=3).\end{cases}
	\end{multline}
These limiting behaviors are indicated on the graphs as blue lines. If $a>1$, $\psi_d(a,b)$ diverges logarithmically for $b\to0$, but in different ways in the two ranges $1<a<2$ and $a>2$. The former range corresponds physically to $0<\mu<\cutoff$, such that the band edge sets the lower cutoff for the pairing interaction, while the latter range is the usual regime, where the Fermi energy is larger than the Debye energy. In two dimensions, the asymptotic behavior is quite simple: If $a>2$, we have the well-known result,
	\begin{equation}\label{eq:psi2Damore2}
		\psi_2(a>2,b\to 0)=\int_{-\frac{1}{b}}^{\frac{1}{b}}dx\,\frac{\tanh(x/2)}{2x}
		=\ln\left(\frac{2e^{\gamma}}{\pi b}\right),
	\end{equation}
with $\gamma\approx 0.577$ the Euler constant. The function is independent of $a$, because the DOS is constant over the range of integration when $\mu>\cutoff$. If $1<a<2$, we evaluate the function by extending the integral to reproduce the case $a>2$ and subtracting the difference:
	\begin{multline}\label{eq:psi2Da12}
		\psi_2(1<a<2,b\to 0)=\int_{-\frac{1}{b}}^{\frac{1}{b}}dx\,\frac{\tanh(x/2)}{2x}
		-\int_{-\frac{1}{b}}^{\frac{1-a}{b}}dx\,\frac{-1}{2x}\\
		=\ln\,\left(\sqrt{a-1}\,\frac{2e^{\gamma}}{\pi b}\right).
	\end{multline}
Equations~(\ref{eq:psi2Damore2}) and (\ref{eq:psi2Da12}) are represented in Fig.~\ref{fig:psi}(top) as red lines. In order to obtain the exact asymptotic behavior for $b\to0$ in three dimensions, we introduce the function $t(x)$ as a piece-wise linear approximation of $\tanh(x/2)$---namely, $-1$ for $x<-2$, $x/2$ for $|x|<2$, and $+1$ for $x>2$---and we calculate analytically the integral with $\tanh(x/2)$ replaced by $t(x)$. The difference between the latter approximation and the exact result is
	\begin{multline*}
		\lim_{b\to0}\int_{\frac{1-\min(a,2)}{b}}^{\frac{1}{b}}dx\,
		\sqrt{bx+a-1}\,\frac{\tanh(x/2)-t(x)}{2x}\\
		=\sqrt{a-1}\ln\left(\frac{4e^{\gamma-1}}{\pi}\right).
	\end{multline*}
Expanding the nonsingular terms to leading order in $b$, we finally get in the regime $0<\mu<\cutoff$:
	\begin{multline}\label{eq:psi3Da12}
		\psi_3(1<a<2,b\to 0)\\=\sqrt{a}+\sqrt{a-1}\,\ln\left(\frac{a-1}{\sqrt{a}+\sqrt{a-1}}
		\frac{8e^{\gamma-2}}{\pi b}\right),
	\end{multline}
and in the regime $\mu>\cutoff$:
	\begin{multline}\label{eq:psi3Damore2}
		\psi_3(a>2,b\to 0)=\sqrt{a}+\sqrt{a-2}+\sqrt{a-1}\\
		\times\ln\left(\frac{a-1}{\sqrt{a}+\sqrt{a-1}}
		\sqrt{\frac{\sqrt{a-1}-\sqrt{a-2}}{\sqrt{a-1}+\sqrt{a-2}}}
		\frac{8e^{\gamma-2}}{\pi b}\right).
	\end{multline}
These asymptotic behaviors are indicated in Fig.~\ref{fig:psi}(bottom) as red lines. Lastly, in the high-density, high-$T_c$ sector $a>2$ and $b\to\infty$, the function reduces simply to $\psi_d(a,b)=(a-1)^{d/2-1}/(2b)$.

Figure~\ref{fig:psi} also shows a particular cut at the value $\psi_d=2.5$. Since the BCS equation (\ref{eq:tc-Nbandsa}) for one band is simply $\psi_d=1/\coupling$, these cuts show the locus of the solutions $(a,b)=(1+\tilde{\mu},\tilde{T}_c)$ for $\coupling=0.4$. Note that the approximations (\ref{eq:psi2Damore2}) to (\ref{eq:psi3Damore2}) shown in red underestimate the function $\psi_d$ at low $b$; using them instead of the exact functions thus leads to underestimating $T_c$.

\section{One parabolic band in 2D and 3D}
\label{sec:1band}

\subsection{Analytical results}

For a single band, we place the origin of energy at the bottom of the band and we use the band mass as the reference mass. The coupled equations (\ref{eq:tc-Nbands}) for $T_c$ become simply:
	\begin{equation}\label{eq:tc3D}
		1=\coupling\,\psi_d\left(1+\tilde{\mu},\tilde{T}_c\right),\quad
		\tilde{n}=-\tilde{T}_c^{\frac{d}{2}}\,\text{Li}_{\frac{d}{2}}
		\left(-e^{\tilde{\mu}/\tilde{T}_c}\right).
	\end{equation}
In 2D the relation between $\tilde{n}$ and $\tilde{\mu}$ can be trivially inverted and the two equations reduce to a single implicit relation for $\tilde{T}_c$ as a function of $\tilde{n}$ and $\coupling$:
	\begin{equation}\label{eq:tc2D}
		1=\coupling\,\psi_2\left(1+\tilde{T}_c\ln
		\big(e^{\tilde{n}/\tilde{T}_c}-1\big),\tilde{T}_c\right).
	\end{equation}
At not too low density, we see from the asymptotic expressions indicated in Fig.~\ref{fig:psi} that the pairing temperature crosses over between two regimes at $\tilde{\mu}=1$. In 2D we have
	\begin{equation}\label{eq:tc2D-highn}
		\tilde{T}_c\approx\frac{2e^{\gamma}}{\pi}\exp\left(-\frac{1}{\coupling}\right)
		\times\begin{cases}\sqrt{\tilde{\mu}} & \tilde{\mu}\lesssim1\\[1em]
			1 & \tilde{\mu}>1 \end{cases}\qquad(d=2).
	\end{equation}
$\tilde{T}_c$ is independent of $\tilde{\mu}$ (hence of $\tilde{n}$) for $\tilde{\mu}>1$, due to the constant DOS and the conventional BCS expression is recovered. In 3D we find
	\begin{multline}\label{eq:tc3D-highn}
		\tilde{T}_c\approx\frac{8e^{\gamma-2}}{\pi}\,\sqrt{\tilde{\mu}}\,
		\exp\left(-\frac{1}{\coupling\sqrt{\tilde{\mu}}}\right)
		\frac{e^{\sqrt{1+1/\tilde{\mu}}}}{1+\sqrt{1+1/\tilde{\mu}}}\\
		\times\begin{cases}
		1 &
		\tilde{\mu}\lesssim1\\[1em]
		e^{\sqrt{1-1/\tilde{\mu}}}\,\sqrt{\frac{1-\sqrt{1-1/\tilde{\mu}}}
		{1+\sqrt{1-1/\tilde{\mu}}}} & \tilde{\mu}>1
		\end{cases}
		\qquad(d=3).
	\end{multline}
The product $\coupling\sqrt{\tilde{\mu}}$ is the coupling evaluated at the chemical potential, which enters the exponential as expected. Equations (\ref{eq:tc2D-highn}) and (\ref{eq:tc3D-highn}) are identical to Eqs.~(2) and (3) of Ref.~\onlinecite{Eagles-1967} if we admit that $k_{\text{B}}T_c=(e^{\gamma}/\pi)\Delta$, which is true in the regime of validity of these expressions, but not in the low-density and/or strong-coupling regimes (see below). We emphasize that these approximations result from expanding the function $\psi_d(a,b)$ in the limit $b\to 0$ for $a>1$ and are therefore accurate only in the limit $T_c\to 0$ at finite positive $\mu$. These equations are \emph{not} accurate in the high-density regime where $T_c$ is large. Our numerical results show that Eqs.~(\ref{eq:tc2D-highn}) and (\ref{eq:tc3D-highn}) provide a rather poor approximation as soon as $T_c$ reaches a few tenths of $\cutoff$.

We now turn to the low-density region. In 2D any cut of the function $\psi_2(a,b)$ at the value $1/\coupling$ converges at $b=0$ to a value $a<1$, given by the relation $-\ln\sqrt{1-a}=1/\coupling$ (see Fig.~\ref{fig:psi}). Hence the chemical potential converges to a finite negative value $\tilde{\mu}_{\min}=-e^{-2/\coupling}$ when the density approaches zero. This is a conjugated effect of the pairing interaction and the DOS discontinuity: At any finite coupling the momentum distribution is spread and a negative chemical potential leads to a finite density even at zero temperature. The chemical potential at zero density is related to the energy $E_b$ of the two-particle bound state by $\tilde{\mu}_{\min}=\tilde{E}_b/(2+\tilde{E}_b)$. Equation (\ref{eq:tc2D}) for $\tilde{T}_c\to 0$ becomes $1=-\coupling\ln\sqrt{-\tilde{T}_c\ln(e^{\tilde{n}/\tilde{T}_c}-1)}$, which can be solved for $\tilde{n}$ as a function of $\tilde{T}_c$: $\tilde{n}=\tilde{T}_c\ln\{1+\exp[-\exp(-2/\coupling)/\tilde{T}_c]\}$. The latter expression shows that $\tilde{n}$ is smaller than $\tilde{T}_c$ when both approach zero, such that in this limit we can replace $\ln(e^{\tilde{n}/\tilde{T}_c}-1)$ by $\ln(\tilde{n}/\tilde{T}_c)$. We thus find the solution,
	\begin{equation}\label{eq:tc2D-lown}
		\tilde{T}_c=\tilde{n}\,
		\exp\left[W\left(\frac{e^{-2/\coupling}}{\tilde{n}}\right)\right]
		\qquad(d=2,\,n\to0).
	\end{equation}
$W(x)$ is the Lambert function (or ``product logarithm''), which gives the principal solution of the equation $x=We^W$. Equation (\ref{eq:tc2D-lown}) is nonanalytic in both $\coupling$ and $\tilde{n}$. It gives a $T_c$ starting with an infinite slope at $n=0$ and increasing faster than any power of $n$ (in the sense that the running exponent given by the logarithmic derivative approaches zero for $n\to0$). An approximation of (\ref{eq:tc2D-lown}) valid to logarithmic accuracy was given earlier \cite{[{See }][{, and references therein.}] Chubukov-2016}.

In 3D the function $\psi_3(a<1,0)$ approaches $1$ for $a\to1$. Therefore we have the same situation as in 2D if $\coupling>1$. In this case the chemical potential approaches a finite negative value given by the solution of $\sqrt{\tilde{\mu}_{\min}+1}-\sqrt{-\tilde{\mu}_{\min}}\sin^{-1}\big(\sqrt{\tilde{\mu}_{\min}+1}\big)=1/\coupling$ as the density approaches zero. Since $\tilde{\mu}$ is finite and negative in the limit $\tilde{T}_c\to0$, we can use the asymptotic expression $-\text{Li}_{3/2}(-e^x)\to e^x$ for $x\to-\infty$ and get the chemical potential $\tilde{\mu}=\tilde{T}_c\ln(\tilde{n}/\tilde{T}_c^{3/2})$. Equation (\ref{eq:tc3D}) can then be solved for $\tilde{T}_c$ in the relevant regime $-\tilde{\mu}\ll1$, by making use of Eq.~(\ref{eq:psialess1}) to leading order in $1-a$. This yields
	\begin{multline}\label{eq:tc3D-lown-strong}
		\tilde{T}_c\approx\tilde{n}^{\frac{2}{3}}\,\exp\left[W\left(\frac{8(1/\coupling-1)^2}
		{3\pi^2\tilde{n}^{\frac{2}{3}}}\right)\right]\\
		\qquad(d=3,\,\coupling>1,\,n\to0).
	\end{multline}
Like in 2D, $T_c$ starts with an infinite slope at $n=0$ and increases faster than any power of $n$ if $\coupling>1$. If $\coupling<1$ there is no finite solution $a$ to the equation $\psi_3(a,0)=1/\coupling$, meaning that $\mu=0$ at zero density. As can be seen in Fig.~\ref{fig:psi}, the curvature along the cut for $\psi_3>1$ is such that $\tilde{\mu}>\tilde{T}_c$. In the limit $\tilde{T_c}\to0$ we can use the large-$x$ expansion $-\text{Li}_{3/2}(-e^x)\to4/(3\sqrt{\pi})x^{3/2}$ and recover $\tilde{\mu}=(3\sqrt{\pi}\tilde{n}/4)^{2/3}$, which is the zero-temperature noninteracting result. Using the asymptotic form (\ref{eq:psi3Da12}) we finally obtain
	\begin{multline}\label{eq:tc3D-lown-weak}
		\tilde{T}_c=\frac{8e^{\gamma-2}}{\pi}
		\left(\frac{3\sqrt{\pi}\tilde{n}}{4}\right)^{\frac{2}{3}}
		\exp\left[-\left(\frac{1}{\coupling}-1\right)
		\left(\frac{4}{3\sqrt{\pi}\tilde{n}}\right)^{\frac{1}{3}}\right]\\
		\qquad(d=3,\,\coupling<1,\,n\to0).
	\end{multline}
This function starts with \emph{zero} slope at $n=0$ and increases \emph{slower} than any power of $n$. It is exactly equivalent to the result \cite{Leggett-1980} $k_{\mathrm{B}}T_c=(8e^{\gamma-2}/\pi)E_{\mathrm{F}}\exp[\pi/(2k_{\mathrm{F}}a_s)]$ if the $s$-wave scattering length $a_s$ is computed with our interaction potential, namely $4\pi\hbar^2a_s/m=V/(\bar{\lambda}-1)$. This potential has no bound state for two particles if $\bar{\lambda}<1$, which explains why $\mu=0$ at zero density in this case. The change of behavior at $\coupling=1$ is discontinuous according to Eqs.~(\ref{eq:tc3D-lown-strong}) and (\ref{eq:tc3D-lown-weak}), both functions giving $\tilde{T}_c\propto\tilde{n}^{2/3}$ with different pre-factors.

The analytical expressions (\ref{eq:tc2D-highn})--(\ref{eq:tc3D-lown-weak}) are compared below with the numerical results. Note that if the reference mass is not the band mass $m_{\alpha}$, one must replace $\tilde{n}$ by $\tilde{n}(m/m_{\alpha})^{d/2}$ in these equations.

As the BCS mean-field theory is not believed to be a useful model in 1D, we have not discussed this case. For completeness, and because it has been argued that the singularity of the 1D DOS could induce large enhancements of $T_c$ in striped quasi-1D superconductors \cite{Perali-1996}, we show in Appendix~\ref{app:1D} that the pairing temperature is also continuous and nonanalytic at the bottom of a 1D band.

Before closing this section, we point out that the solution of the gap equation at $T=0$ does not generally allow one to deduce $T_c$. Although the focus of the present paper is on $T_c$, we give in Appendix~\ref{app:ratio} exact results for the zero-temperature gap in 2D at low density, for the purpose of showing that the usual BCS gap to $T_c$ ratio is not obeyed in this limit.

\subsection{Numerical results}

\begin{figure}[tb]
\includegraphics[width=0.7\columnwidth]{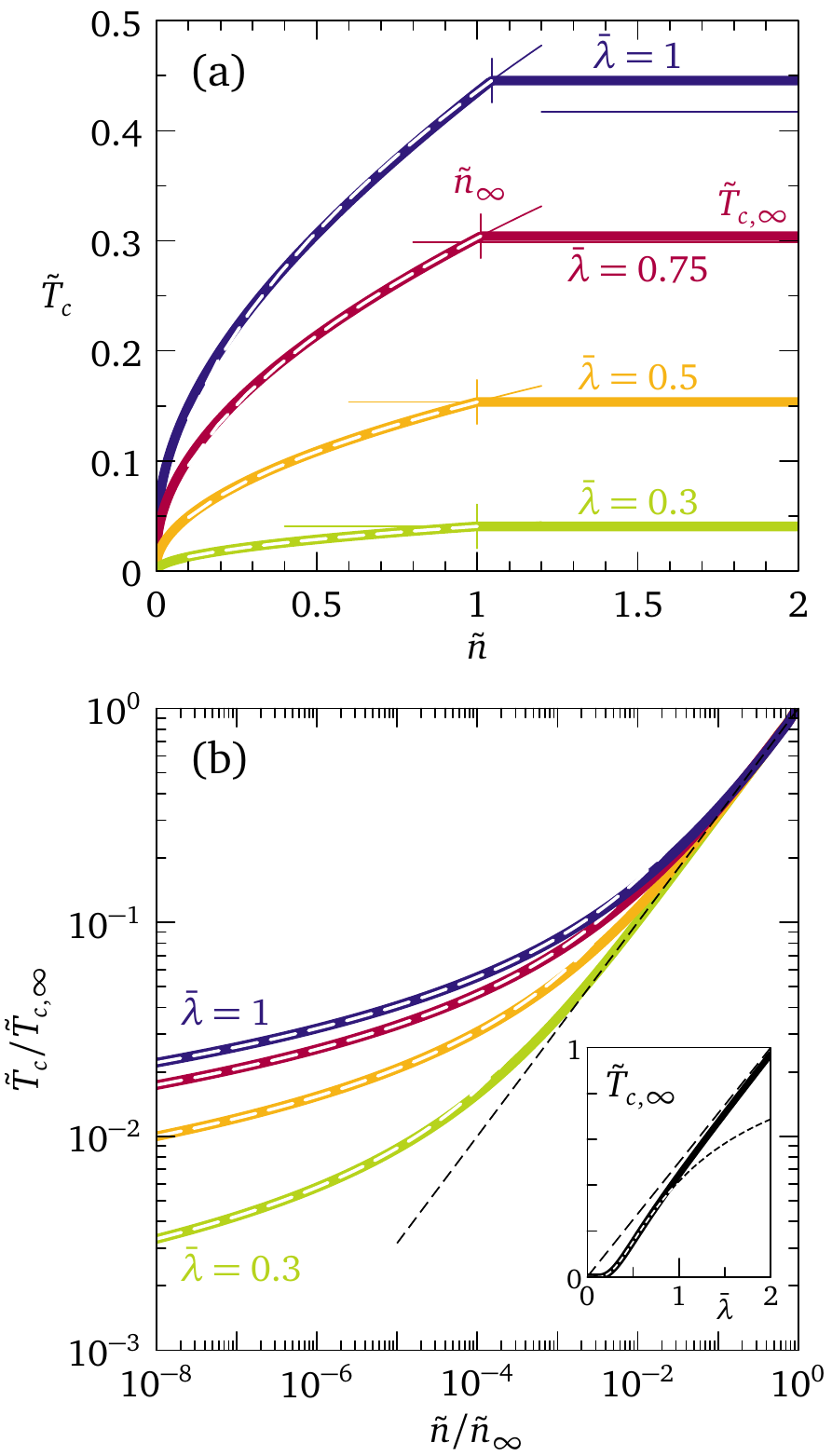}
\caption{\label{fig:1band-2D}
(a) Pairing temperature as a function of electron density for one parabolic band in two dimensions. $T_c$ is expressed in units of $\cutoff/k_{\mathrm{B}}$ and $n$ in units of $m\frequency/(\pi\hbar)$. The thin horizontal lines show the approximate solution (\ref{eq:tc2D-highn}) for each $\coupling$. The vertical bars indicate $\tilde{n}_{\infty}$ [Eq.~(\ref{eq:ninf})]. The dashed lines show the approximate scaling $\tilde{T}_c=\tilde{T}_{c,\infty}(\tilde{n}/\tilde{n}_{\infty})^{1/2}$. (b) Same data normalized. The white dashed lines are the prediction of Eq.~(\ref{eq:tc2D-lown}) and the black dashed line indicates the square-root behavior for $\tilde{n}\lesssim\tilde{n}_{\infty}$. (Inset) Maximum pairing temperature as a function of $\coupling$ (solid line), compared with Eq.~(\ref{eq:tc2D-highn}) (dotted) and $\coupling/2$ (dashed).
}
\end{figure}

\begin{figure}[tb]
\includegraphics[width=0.7\columnwidth]{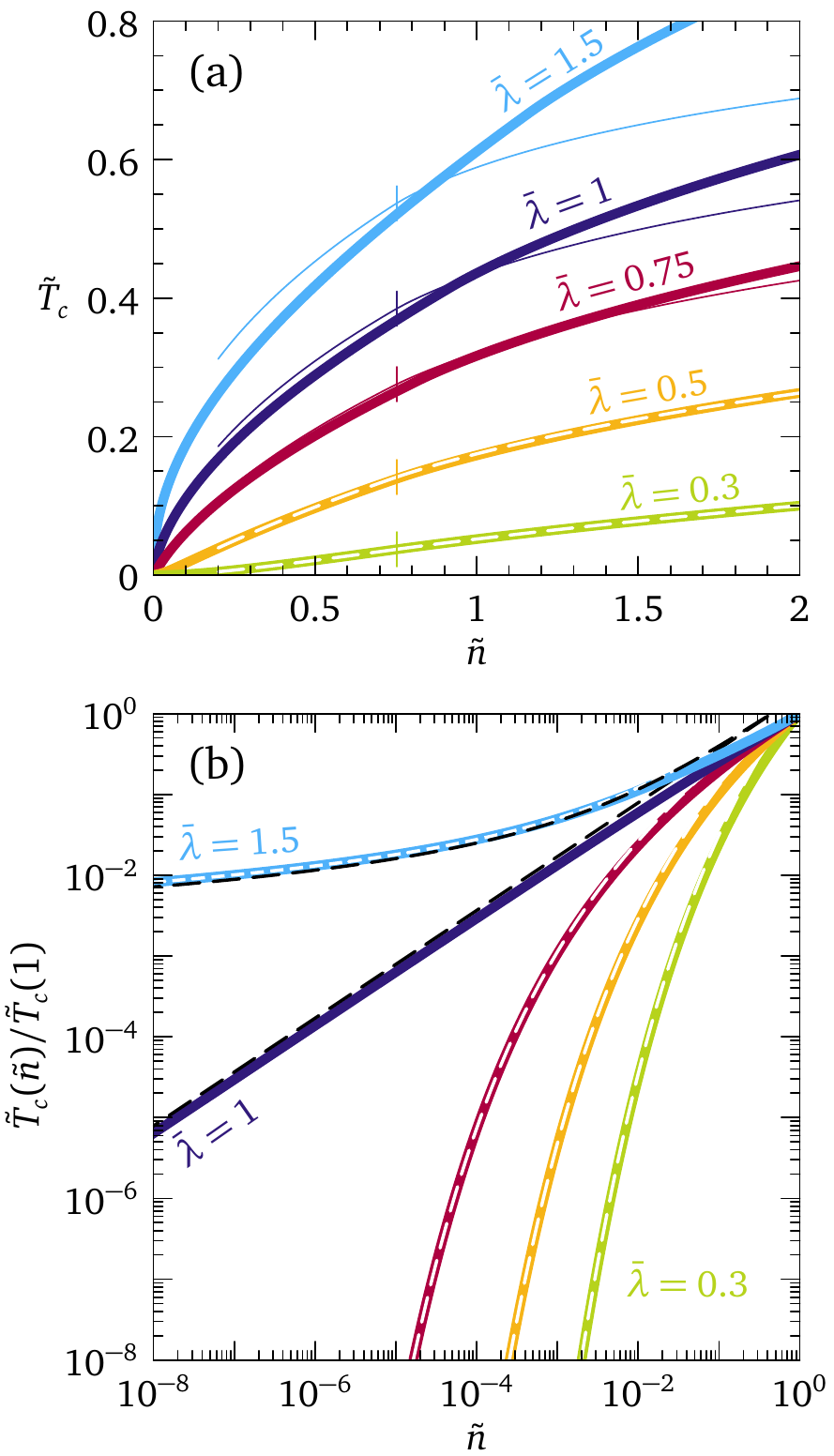}
\caption{\label{fig:1band-3D}
(a) Pairing temperature as a function of electron density for one parabolic band in three dimensions. $T_c$ is expressed in units of $\cutoff/k_{\mathrm{B}}$ and $n$ in units of $2[m\frequency/(2\pi\hbar)]^{3/2}$. The thin and dashed lines show Eq.~(\ref{eq:tc3D-highn}), evaluated using $\tilde{\mu}_0=(3\sqrt{\pi}\tilde{n}/4)^{2/3}$ for $\tilde{\mu}$. The vertical bars indicate $\tilde{\mu}_0=1$. (b) Same data on a log-log scale. The dashed lines show Eqs.~(\ref{eq:tc3D-lown-strong}) and (\ref{eq:tc3D-lown-weak}). Equation~(\ref{eq:tc3D-lown-weak}) was used for $\coupling=1$. The short-dashed white line for $\coupling=1.5$ is obtained without expanding Eq.~(\ref{eq:psialess1}) around $a=1$ (see text).
}
\end{figure}

The numerical solution of Eq.~(\ref{eq:tc2D}) is shown in Fig.~\ref{fig:1band-2D}. $\tilde{T}_c$ reaches a plateau at high density due to the constant DOS of the band. For $\coupling$ of order one, the value $T_{c,\infty}$ on the plateau departs significantly from the approximate solution (\ref{eq:tc2D-highn}), which becomes worse with increasing $\coupling$, while the simple large-$T_c$ result $\tilde{T}_{c,\infty}=\coupling/2$ becomes increasingly reliable [inset of Fig.~\ref{fig:1band-2D}(b)]. The density $\tilde{n}_{\infty}$ at which the plateau is reached corresponds to $\mu-\cutoff$ coinciding with the bottom of the band, which means:
	\begin{equation}\label{eq:ninf}
		\tilde{n}_{\infty}=\tilde{T}_{c,\infty}\ln\left(e^{1/\tilde{T}_{c,\infty}}+1\right).
	\end{equation}
For $\tilde{n}\lesssim\tilde{n}_{\infty}$, Eq.~(\ref{eq:tc2D-highn}) gives $\tilde{T}_c/\tilde{T}_{c,\infty}\approx\tilde{\mu}^{1/2}$. Since $\tilde{\mu}$ is very close to a linear function of $\tilde{n}$ at intermediate and high densities (see below), we expect to have the universal scaling $\tilde{T}_c/\tilde{T}_{c,\infty}\approx(\tilde{n}/\tilde{n}_{\infty})^{1/2}$. This is well obeyed by the data.

Close to $\tilde{n}=0$ the behavior is nonuniversal, in the sense that the curves do not collapse if $\tilde{n}$ and $\tilde{T}_c$ are rescaled by $\tilde{n}_{\infty}$ and $\tilde{T}_{c,\infty}$ [Fig.~\ref{fig:1band-2D}(b)]. The numerical data are in perfect agreement with the limiting behavior (\ref{eq:tc2D-lown}) at all couplings. The flattening of the curves in the log-log plot shows that the running exponent $\eta(n)$ in $T_c\propto n^{\eta(n)}$ approaches zero for $n\to0$. This is suggestive of a discontinuity in $T_c(n)$ at $n=0$, reminiscent of the DOS discontinuity. However, since Eq.~(\ref{eq:tc2D-lown}) vanishes continuously for $\tilde{n}\to0$, the correct picture is that of a $T_c$ tending asymptotically to a discontinuity of size zero with decreasing $n$.

The numerical results for the 3D case are displayed in Fig.~\ref{fig:1band-3D}. Also shown is the high-density approximation (\ref{eq:tc3D-highn}), evaluated with $\tilde{\mu}$ replaced by its zero-temperature noninteracting value $\tilde{\mu}_0$. The approximation falls on top of the numerical data for small $\coupling$, but deviates significantly for larger coupling. The good agreement at weak coupling is due to a cancellation of errors: the agreement worsens if $\tilde{\mu}$ rather than $\tilde{\mu}_0$ is used in Eq.~(\ref{eq:tc3D-highn}). The reason is that $\mu(T_c)<\mu_0$ and the use of $\mu_0$ always leads to overestimating $T_c$. This happens to compensate the underestimation of $T_c$ due to the use of Eqs.~(\ref{eq:psi3Da12}) and (\ref{eq:psi3Damore2}).

At low density, the change of behavior from a convex increase for $\coupling<1$ to a concave increase for $\coupling>1$ is visible on the log-log plot in Fig.~\ref{fig:1band-3D}(b)---where a convex function has a slope larger than unity. The low-density, low-coupling limit (\ref{eq:tc3D-lown-weak}) describes the numerical data perfectly. The low-density, high-coupling expression (\ref{eq:tc3D-lown-strong}) deviates slightly due to the use of Eq.~(\ref{eq:psialess1}) at lowest order in $1-a$. This small discrepancy disappears if $\tilde{\mu}$ is evaluated without expanding Eq.~(\ref{eq:psialess1}). The value $\coupling=1$ is somewhat peculiar: The numerics shows the expected $\tilde{n}^{2/3}$ scaling, but the pre-factor is neither unity as implied by Eq.~(\ref{eq:tc3D-lown-strong}), nor $0.742$ as given by Eq.~(\ref{eq:tc3D-lown-weak}), but $\sim0.6$.

\begin{figure}[tb]
\includegraphics[width=\columnwidth]{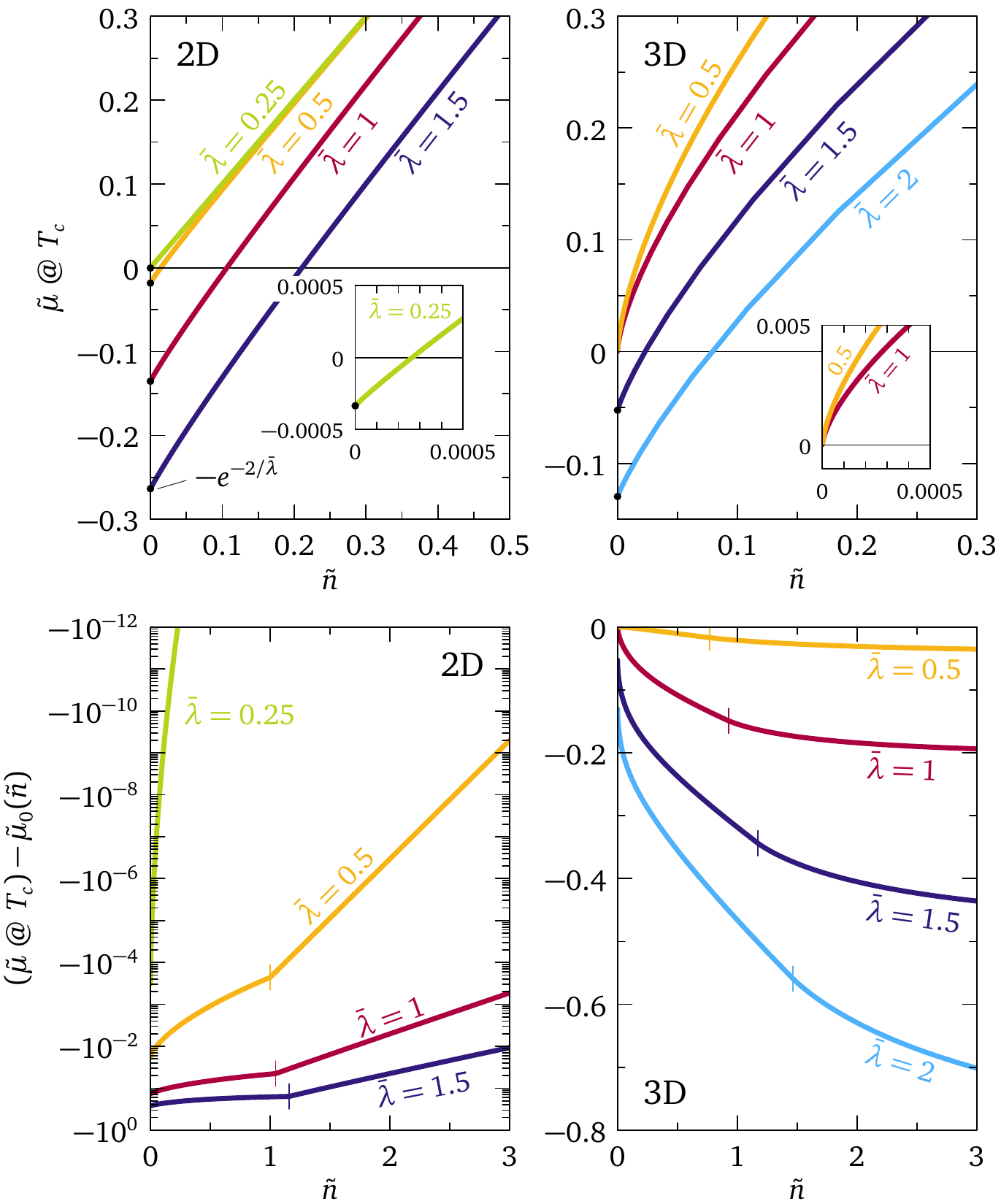}
\caption{\label{fig:1band-mu}
(Top row) Chemical potential at $T_c$ in the low-density limit. The dots show the solution of $\psi_d(1+\tilde{\mu},0)=1/\coupling$, with $\psi_d(a,0)$ given by Eq.~(\ref{eq:psialess1}). The insets show that $\mu(n=0)<0$ in 2D for all $\coupling$, while $\mu(n=0)=0$ in 3D for $\coupling\leqslant1$. (Bottom row) Difference between the chemical potential at $T_c$ and the zero-temperature noninteracting value $\tilde{\mu}_0$. The vertical bars indicate $\tilde{\mu}=1$.
}
\end{figure}

Figure~\ref{fig:1band-mu} shows the chemical potential calculated numerically at $T_c$. In 2D $\mu$ converges to a negative value for any coupling, as discussed above. In 3D $\mu$ tends to zero at $n=0$ if $\coupling\leqslant1$. If $\coupling>1$ it converges to a negative value. The density at which $\mu=0$ is given for $\bar{\lambda}\gtrsim1$ by $\tilde{n}\approx0.62(1-1/\bar{\lambda})^3$. This coincides with the condition $1/(k_{\mathrm{F}}a_s)\approx0.68$. The effect of increasing the pairing interaction is mainly to shift the $\mu(n)$ curve downwards. At $n=0$ this shift is entirely due to the interaction-induced spreading of the momentum distribution. At finite $n$ part of the shift is due to the thermal smearing.

As the density increases, the behavior is qualitatively different in 2D and 3D: while $\tilde{\mu}$ approaches $\tilde{\mu}_0=\tilde{n}$ in 2D, this does not happen in 3D. In 2D the chemical potential is $\tilde{\mu}=\tilde{T}_c\ln(e^{\tilde{n}/\tilde{T}_c}-1)$. Since $\tilde{T}_c$ saturates for $\tilde{n}>\tilde{n}_{\infty}$, we have $\tilde{n}>\tilde{T}_c$ at large $\tilde{n}$ and $\tilde{\mu}$ approaches exponentially the value $\tilde{n}$. This is peculiar to the 2D constant DOS, since both interaction and temperature redistribute states in equal amounts below and above $\mu_0$. In 3D the square-root DOS implies that there are more states added in the tail of the momentum distribution above $\mu_0$, than there are states removed below $\mu_0$. The equilibrium chemical potential must therefore remain below the zero-temperature noninteracting value, by an amount which increases with increasing $\coupling$ and also with increasing $n$.

\section{Multiband effects}
\label{sec:2bands}

The interest raised by multiband superconductors, in particular MgB$_2$ and the iron-based family, has triggered many studies over the years \cite{[{For a recent review, see: }] Zehetmayer-2013}. Here we discuss multiband effects that occur near a band edge and are associated with the low density in one of the bands.

\begin{figure}[tb]
\includegraphics[width=0.9\columnwidth]{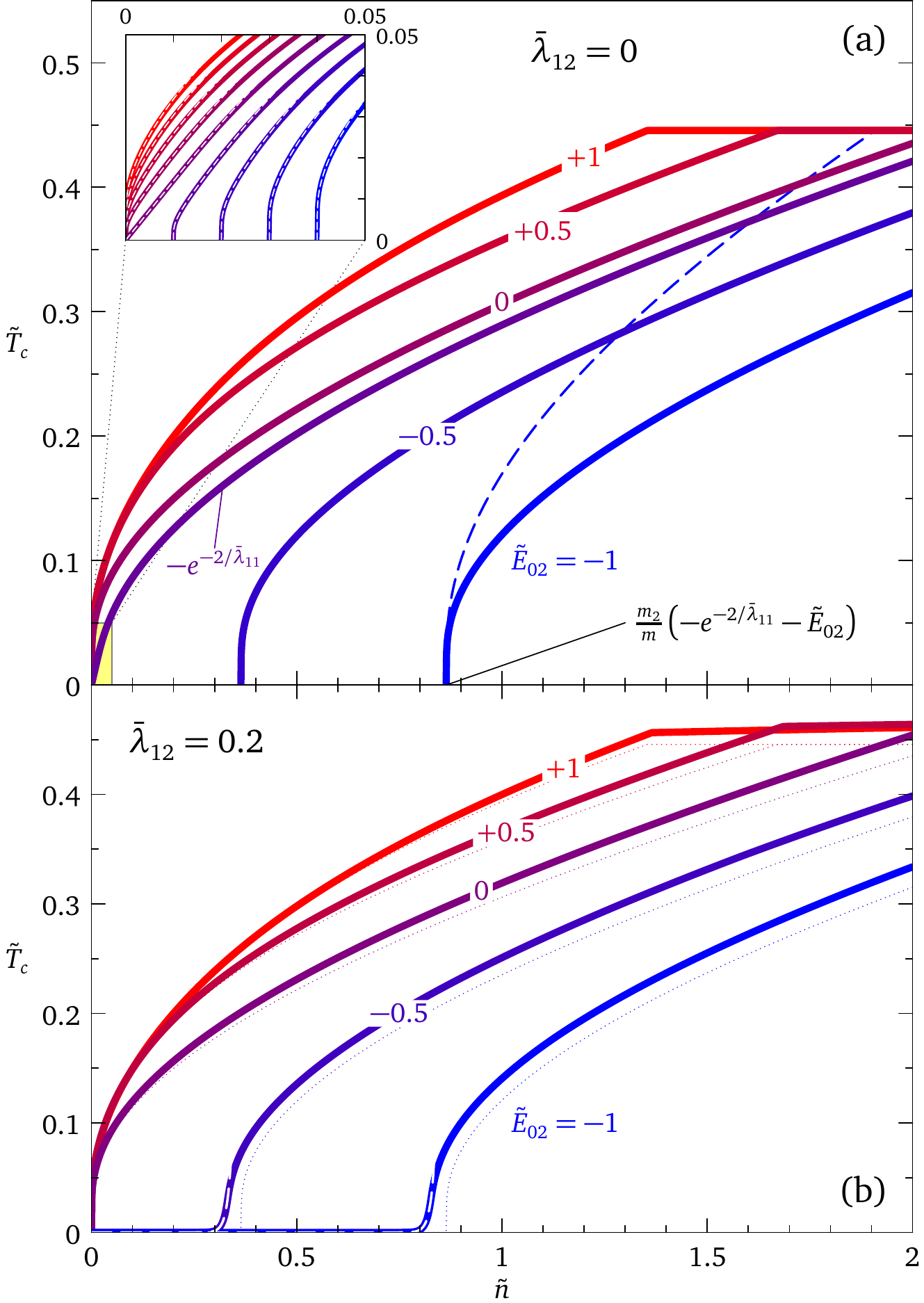}
\caption{\label{fig:2bands-ns}
Pairing temperature for two bands in two dimensions, with couplings $\coupling_{11}=1$, $\coupling_{22}=0$, and (a) $\coupling_{12}=0$, (b) $\coupling_{12}=0.2$. $E_{02}$ is the energy minimum of the second band, measured from the energy minimum of the first. The masses are $m_1=m_2=m$. The dashed line in (a) is the one-band result, shifted horizontally for easier comparison. (Inset) Blowup of the transition region. Curves are shown for $\tilde{E}_{02}=-\exp(-2/\coupling_{11})+\delta\tilde{E}$, with $\delta\tilde{E}$ ranging from $-0.04$ (blue, right) to $+0.04$ (red, left). Note the linear behavior for $\delta\tilde{E}=0$. The white dashed lines show Eq.~(\ref{eq:Tc2bands1}). In (b), the dotted lines are the result for $\coupling_{12}=0$ and the dashed white lines show Eq.~(\ref{eq:Tctail}).
}
\end{figure}

It is clear from the previous section that a knowledge of the self-consistent chemical potential is required to understand the behavior of $T_c$ close to a band minimum. This raises the question of the role played by perturbations that affect the chemical potential, such as the presence of a nonsuperconducting band (NB) beneath the superconducting band (SB). In the absence of interband coupling, the NB can only alter the superconducting properties of the SB by changing the chemical potential. In 2D and in 3D for $\coupling>1$, the key observation was that $\mu$ is finite and negative at the band bottom, such that the nonanalytic behavior of $T_c$ is not controlled by $\mu$. An NB is therefore not expected in general to change this nonanalytic behavior qualitatively. An exception---confirming the rule---occurs when the bottom of the NB coincides precisely with the energy at which the SB begins to be populated. For this peculiar arrangement, the NB controls the relation between $\mu$ and $n$ in the limit $T_c\to0$ and $T_c$ displays a simple analytic dependence on $n$, which is linear in 2D and $\propto n^{2/3}$ in 3D. This is illustrated in Fig.~\ref{fig:2bands-ns}(a) for the 2D case. Solving the coupled equations (\ref{eq:tc-Nbands}) for two bands in the appropriate regime, we get a relation between $\tilde{n}$ and $\tilde{T}_c$ which is accurate near the band minimum:
	\begin{multline}\label{eq:Tc2bands1}
		\tilde{n}=\tilde{T}_c\left[\frac{m_1}{m}
		\ln\left(1+e^{\frac{-\exp(-2/\coupling_{11})}{\tilde{T}_c}}\right)
		\right.\\ \left.
		+\frac{m_2}{m}\ln\left(1+e^{\frac{-\exp(-2/\coupling_{11})-\tilde{E}_{02}}
		{\tilde{T}_c}}\right)\right].
	\end{multline}
This reproduces the near-band edge behavior as shown in Fig.~\ref{fig:2bands-ns}(a) and in particular gives the linear dependence $\tilde{n}=\tilde{T}_c(m_2/m)\ln(2)$ at the transition point where $\tilde{E}_{02}=-\exp(-2/\coupling_{11})$.

If $\tilde{E}_{02}<-\exp(-2/\coupling_{11})$, the SB is not populated at low density and superconductivity appears at some finite density. In all cases, the $T_c(n)$ curve is ``stretched'' to higher densities with respect to the one-band result due to the carriers ``lost'' in the NB. For instance, if the band minima are degenerate, we see from Eqs.~(\ref{eq:tc-Nbands}) that the one-band $T_c(n)$ curve is simply modified by a rescaling of the density $n\to n/[1+(m_1/m_2)^{d/2}]$. This implies that the pairing temperature is necessarily reduced by a nonsuperconducting band, in the absence of interband coupling.

Both attractive and repulsive interband interactions increase $T_c$ for two bands \cite{Suhl-1959, Bussmann-Holder-2004}, as illustrated by the fact that Eq.~(\ref{eq:tc-Nbands}) involve only $\coupling_{12}^2$: interband interactions do not induce interband pairing in the present model, but reinforce the intraband pairing by second-order processes involving the other band. If $T_c$ starts at finite density, the interband coupling leads to a tail in the $T_c(n)$ curve. In the regime where the chemical potential is below the SB but well into the NB, we find, for instance, in 2D:
	\begin{equation}\label{eq:Tctail}
		\tilde{T}_c=\frac{2e^{\gamma}}{\pi}\sqrt{\tilde{n}\frac{m}{m_2}}\exp\left\{\textstyle
		\frac{1}{\coupling_{12}^2}\left[\coupling_{11}+\frac{2}{\ln\left(-\tilde{E}_{02}
		-\tilde{n}\frac{m}{m_2}\right)}\right]\right\}.
	\end{equation}
This is compared with the numerical result in Fig.~\ref{fig:2bands-ns}(b).

\begin{figure}[tb]
\includegraphics[width=\columnwidth]{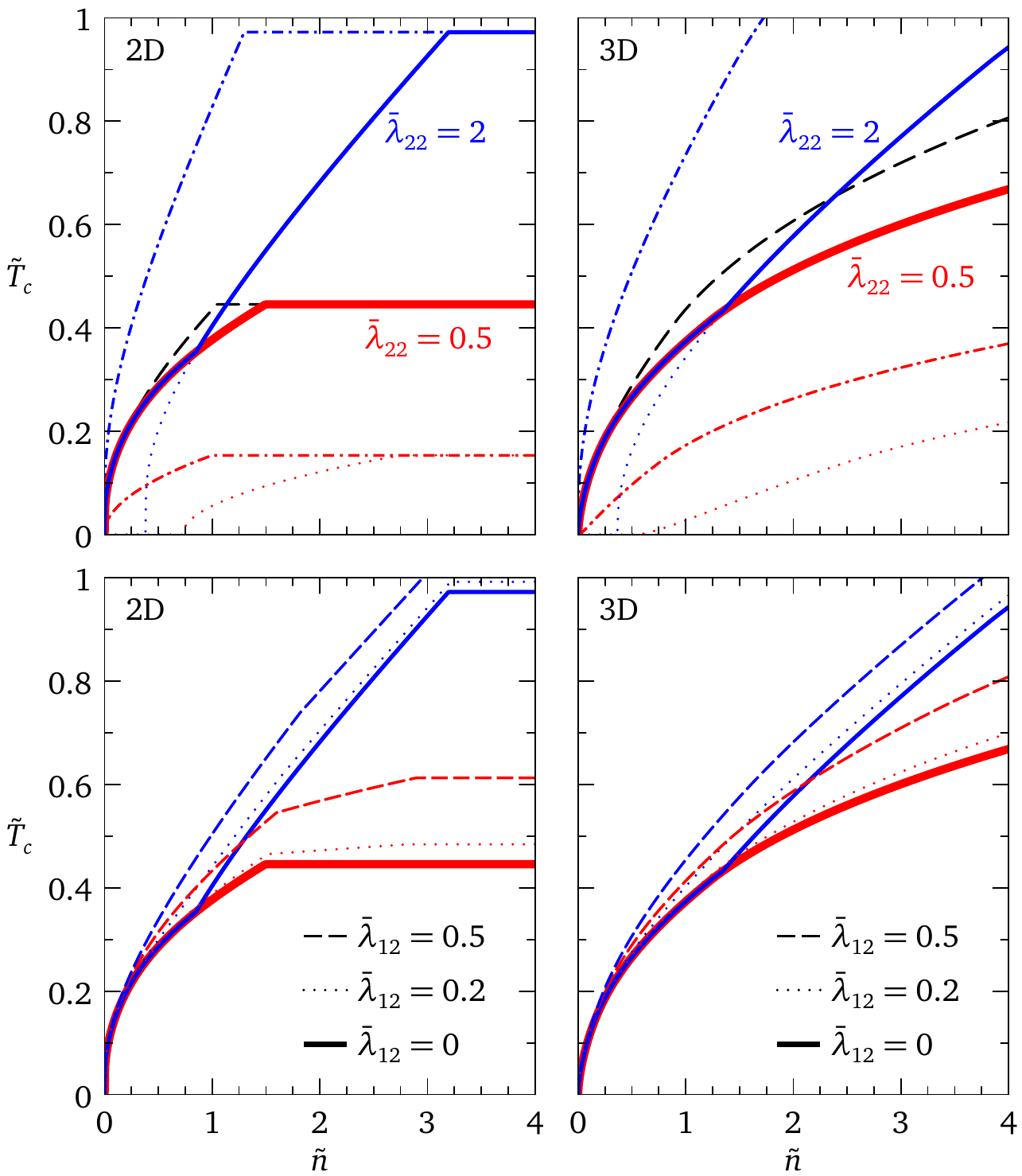}
\caption{\label{fig:2bands-ss}
(Top panels) Change of $T_c$ induced by a second superconducting band in 2D and 3D, without interband coupling. (Dashed black lines) Pairing temperature for a single band with coupling $\coupling_{11}=1$ and mass $m_1/m=1$. (Solid lines) Pairing temperature for the two-band system with $\tilde{E}_{02}-\tilde{E}_{01}=0.75$, $m_2=m_1$, and coupling $\coupling_{22}=0.5$ (red) and $\coupling_{22}=2$ (blue). (Dash-dotted) Case of the second band alone. (Dotted) Case of the two-band system with $\coupling_{11}=0$. (Bottom panels) Increase of $T_c$ by interband coupling. (Solid lines) No interband coupling, same data and coloring as in the top panels. Dotted and dashed lines correspond to interband coupling $\coupling_{12}=0.2$ and $0.5$, respectively.
}
\end{figure}

We move on to the case of two superconducting bands and begin with general trends. The observation that $T_c$ is an increasing function of density \cite{Fernandes-2013} remains true in the near-band edge regime. It is possible to show that the property $dT_c/dn\geqslant0$ is guaranteed by Eq.~(\ref{eq:tc-Nbands}) for an arbitrary number of bands and any values of the coupling constants. Reference~\onlinecite{Lin-2014} reports a nonmonotonic dependence of the pairing temperature on carrier concentration in doped SrTiO$_3$: this can not be interpreted on the basis of Eq.~(\ref{eq:tc-Nbands}) without invoking density-dependent interactions.

A second band can nevertheless lead to a decrease of $T_c$ at fixed density. Specifically, consider a one-band system at some density with coupling $\coupling_{11}$ and pairing temperature $T_c^0$; add a second band at higher energy with coupling $\coupling_{22}\leqslant\coupling_{11}$ and no interband coupling; then the two-band system with the same density has $T_c\leqslant T_c^0$. This can be rigorously proven by manipulating Eqs.~(\ref{eq:tc-Nbands}).

If the second band has a coupling $\coupling_{22}>\coupling_{11}$, $T_c$ exceeds $T_c^0$ at high enough density and follows the dependence that would correspond to a nonsuperconducting first band. These various trends are illustrated in Fig.~\ref{fig:2bands-ss} (top panels). Figure~\ref{fig:2bands-ss} (bottom panels) shows the effect of interband interaction, which is generically an increase of $T_c$.

\section{Conclusion}

In summary, in the low-density regime where the dynamical range of the pairing interaction is set by the band edge, the pairing temperature $T_c$ depends on the electron density $n$ in a nonanalytic way. For parabolic bands, we provided exact asymptotic formulas describing this dependency, taking into account the energy variation of the electronic DOS, as well as the variation of the chemical potential with interaction strength and temperature. In one and two dimensions and in three dimensions at strong enough coupling---in other words, when there is a bound solution to the two-particle problem---the chemical potential (at $T_c$) becomes negative at low density: As a result the $T_c(n)$ curve starts with infinite slope and increases faster than any power of $n$. Otherwise, i.e., in three dimensions at weak coupling, the chemical potential approaches zero at low density, the $T_c(n)$ curve starts with zero slope, and it increases slower than any power of $n$.

Our results may be relevant for low-density superconductors. In SrTiO$_3$, oxygen reduction and niobium doping allows one to tune the carrier density \cite{Lin-2014} in a range such that the dimensionless density $\tilde{n}$ varies typically between $10^{-2}$ and $10$. In the LaAlO$_3$/SrTiO$_3$ interface, the field-effect induced sheet carrier density can also be tuned \cite{Caviglia-2008} such that $\tilde{n}$ varies typically from $10^{-1}$ to $1$. In the low-density range of these domains, our exact formulas differ from the usual formulas valid at higher densities. In the numerical illustrations of the present paper we have used coupling constants $\bar{\lambda}$ of order one, which may appear very large in comparison to the typical values of the order $0.1$ reported for SrTiO$_3$. We emphasize that our definition of the coupling constants differs from the usual definition, such that in three dimensions ours are bigger that the usual ones by a factor $(\mu/\cutoff)^{1/2}$, which is typically three in SrTiO$_3$.

The observation that $T_c$ is a nonanalytic function of $n$ near a band bottom calls for a reconsideration of the problem of shape resonances. These refer to oscillations of $T_c$ in a quasi-two-dimensional superconductor confined in a slab, as a function of the slab thickness. The oscillations arise when the chemical potential crosses the bottom of one of the confinement-induced subbands and were presented in the literature on the subject as discontinuities \cite{Romero-Bermudez-2014}. Our results show that such discontinuities are artifacts, because $T_c$ vanishes continuously at a band edge in any dimension. The actual dependence of $T_c$ on the slab thickness is therefore a continuous function, which remains to be investigated. A particularly interesting system in this respect is the LaAlO$_3$/SrTiO$_3$ interface, which cumulates the characteristics of being a low-density superconducting system, confined in a quasi-two-dimensional geometry, and also a multiband system.

\acknowledgments
We acknowledge fruitful discussions with J.-M. Triscone, S. Gariglio, and M. Grilli. We are grateful to A. V. Chubukov for drawing our attention to a previous approximation of Eq.~(\ref{eq:tc2D-lown}) and to P. Pieri for pointing out the equivalence of Eq.~(\ref{eq:tc3D-lown-weak}) with a known result in the BEC limit. This work was supported by the Swiss National Science Foundation under Division II.

\appendix

\section{Simple proof of Eq.~(\ref{eq:density})}
\label{app:density}

The density of a free-electron gas in dimension $d$ is proportional to the volume of the $d$-dimensional Fermi sphere, smeared by the Fermi function:
	\begin{align*}
		n&=2\int\frac{d^dk}{(2\pi)^d}\frac{1}{\exp\left(
		\frac{\hbar^2k^2/2m-\mu}{k_{\text{B}}T}\right)+1}\\
		&=2\int\frac{d^dk}{(2\pi)^d}\frac{x}{x-\exp\left(
		\frac{\hbar^2k^2}{2mk_{\text{B}}T}\right)},
	\end{align*}
with $x=-\exp\left(\mu/k_{\text{B}}T\right)$. In order to evaluate the integral, we use the expansion $x/(x-a)=-\sum_{q=1}^{\infty}x^q/a^q$ and we write $k^2=\sum_{i=1}^dk_i^2$. This leads to a product of Gaussian integrals:
	\begin{align*}
		n&=-2\sum_{q=1}^{\infty}x^q\prod_{i=1}^d\int_{-\infty}^{\infty}
		\frac{dk_i}{2\pi}\exp\left(-q\frac{\hbar^2k_i^2}{2mk_{\text{B}}T}\right)\\
		&=-2\sum_{q=1}^{\infty}x^q\prod_{i=1}^d
		\sqrt{\frac{mk_{\text{B}}T}{2\pi\hbar^2q}}
		=-2\left(\frac{mk_{\text{B}}T}{2\pi\hbar^2}\right)^{\frac{d}{2}}
		\sum_{q=1}^{\infty}\frac{x^q}{q^{d/2}}.
	\end{align*}
Considering the Taylor expansion of the polylogarithm, we see that the $q$-sum in the last expression is $\text{Li}_{d/2}(x)$, which proves Eq.~(\ref{eq:density}).

\section{Results for one parabolic band in 1D}
\label{app:1D}

\begin{figure}[tb]
\includegraphics[width=0.7\columnwidth]{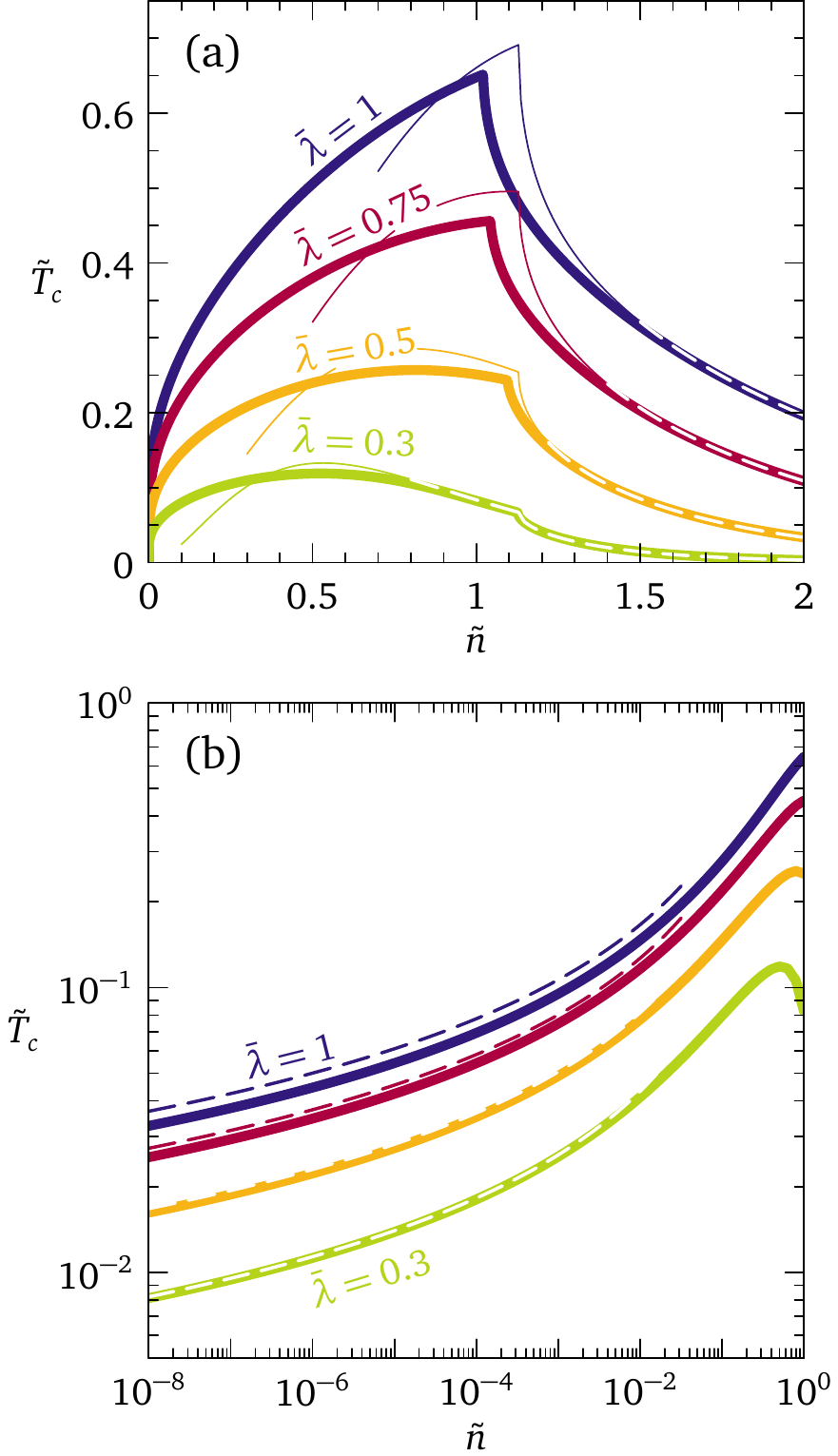}
\caption{\label{fig:1band-1D}
(a) Pairing temperature as a function of electron density for one parabolic band in one dimension. $T_c$ is expressed in units of $\cutoff/k_{\mathrm{B}}$ and $n$ in units of $2[m\frequency/(2\pi\hbar)]^{1/2}$. The thin and dashed lines show Eq.~(\ref{eq:tc1D-highn}), evaluated using $\tilde{\mu}_0=(\pi/4)\tilde{n}^2$ for $\tilde{\mu}$. (b) Same data on a log-log scale. The dashed lines show Eq.~(\ref{eq:tc1D-lown}).
}
\end{figure}

With the proviso that the factor $(d-1)\pi$ in Eqs.~(\ref{eq:DOS_d}) and (\ref{eq:lambda}) be replaced by 1, Eqs.~(\ref{eq:general})--(\ref{eq:tc-Nbands}) are valid for $d=1$. The asymptotic properties of the function $\psi_1(a,b)$ are
	\begin{align*}
		\psi_1(0<a<1,b\to 0)&=\frac{\sin^{-1}(\sqrt{a})}{\sqrt{1-a}}
		\approx\frac{\pi/2}{\sqrt{1-a}}-1\\
		\psi_1(1<a<2,b\to 0)&=\frac{1}{\sqrt{a-1}}\ln\,\left(
		\frac{a-1}{\sqrt{a}+\sqrt{a-1}}\,\frac{8e^{\gamma}}{\pi b}\right)\\
		\psi_1(a>2,b\to 0)&=\frac{1}{\sqrt{a-1}}\\[-5mm]
		\\ &\hspace*{-2cm}\times\ln\,\left(
		\frac{a-1}{(\sqrt{a}+\sqrt{a-1})(\sqrt{a-1}+\sqrt{a-2})}\,
		\frac{8e^{\gamma}}{\pi b}\right).
	\end{align*}
The corresponding weak-coupling approximations in the regime $\tilde{n}\gtrsim1$ are
	\begin{multline}\label{eq:tc1D-highn}
		\tilde{T}_c\approx\frac{8e^{\gamma}}{\pi}\,
		\exp\left(-\frac{1}{\coupling/\sqrt{\tilde{\mu}}}\right)
		\frac{1}{1+\sqrt{1+1/\tilde{\mu}}}\\
		\times\begin{cases}
		\sqrt{\tilde{\mu}} &
		\tilde{\mu}\lesssim1\\[1em]
		\frac{1}{1+\sqrt{1-1/\tilde{\mu}}} & \tilde{\mu}>1
		\end{cases}
		\qquad(d=1),
	\end{multline}
in agreement with the result of Ref.~\onlinecite{Eagles-1967}, where $\coupling/\sqrt{\tilde{\mu}}$ is the coupling constant evaluated at the chemical potential. In the low-density limit the chemical potential approaches a finite negative value given by the solution of $1/\coupling=\sin^{-1}\big(\sqrt{1+\tilde{\mu}_{\min}}\big)/\sqrt{-\tilde{\mu}_{\min}}\approx(\pi/2)/\sqrt{-\tilde{\mu}_{\min}}-1$. We can use the expansion $-\text{Li}_{1/2}(-e^x)\to e^x$ for large negative $x$ and deduce the relation $\tilde{\mu}=\tilde{T}_c\ln(\tilde{n}/\tilde{T}_c^{1/2})$. We find for $\tilde{T}_c$
	\begin{equation}\label{eq:tc1D-lown}
		\tilde{T}_c\approx\tilde{n}^2\,\exp\left[W\left({\textstyle
		\frac{1}{2}\left(\frac{\pi\coupling}{1+\coupling}\right)^2
		\frac{1}{\tilde{n}^2}}\right)\right]
		\quad(d=1,\,n\to0).
	\end{equation}
The numerical results are shown in Fig.~\ref{fig:1band-1D} and compared with the analytical formulas (\ref{eq:tc1D-highn}) and (\ref{eq:tc1D-lown}). Equation~(\ref{eq:tc1D-highn}) works well at high density, but severely breaks down at low density, even at weak coupling. The cancellation of errors observed in the 3D case also occurs here to some extent, but the main issue is that Eq.~(\ref{eq:tc1D-highn}) fails to describe the regime where $\tilde{\mu}<0$ at low density, while in 3D this regime is absent for $\coupling<1$. Equation~(\ref{eq:tc1D-lown}) is accurate at weak coupling where $-\tilde{\mu}\ll1$ and has the small inaccuracy associated with the approximation made in solving for $\tilde{\mu}$ at larger coupling.

\section{\boldmath Gap to $T_c$ ratio in the low-density limit}
\label{app:ratio}

\begin{figure}[b]
\includegraphics[width=0.75\columnwidth]{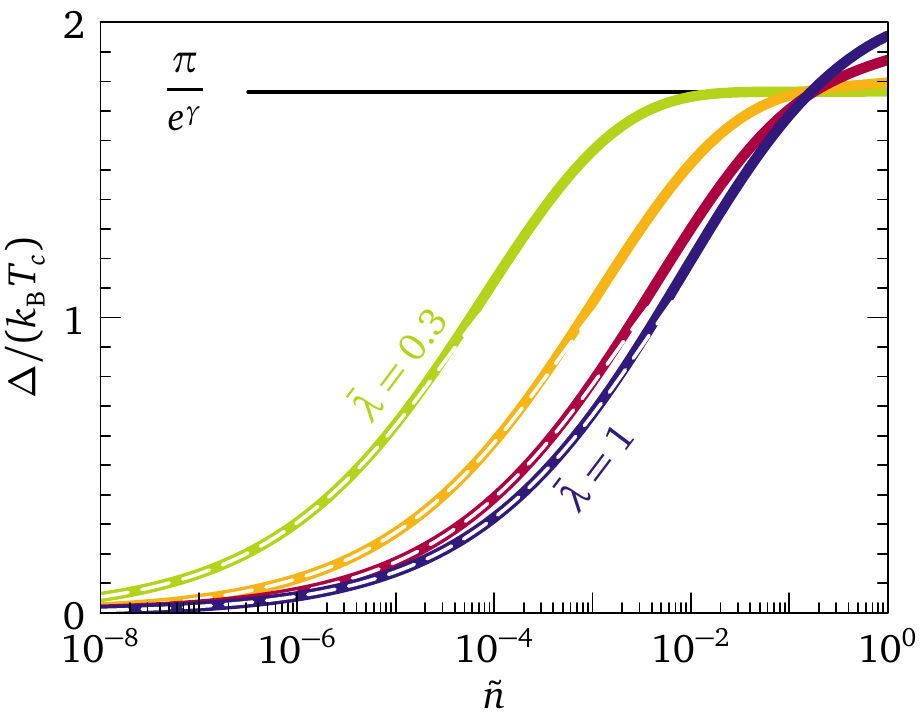}
\caption{\label{fig:ratio-2D}
Zero-temperature gap $\Delta$ to $T_c$ ratio calculated numerically for one parabolic band in two dimensions. Curves are drawn as a function of density $n$ expressed in units of $m\frequency/(\pi\hbar)$ for $\coupling=0.3$, 0.5, 0.75, and 1. The dashed lines show the ratio of Eqs.~(\ref{eq:Deltalown}) and (\ref{eq:tc2D-lown}). The horizontal line indicates the BCS weak-coupling ratio $\pi/e^{\gamma}\approx1.76$.
}
\end{figure}

We give here the zero-temperature gap explicitly for one band in 2D, as a function of density and coupling. This can be combined with the result (\ref{eq:tc2D-lown}) in order to obtain the exact gap to $T_c$ ratio in the low-density limit. The expression of the density at $T=0$ is
	\begin{equation}
		n=\int_{-\infty}^{\infty}d\xi\,N_0(\mu+\xi)\left(1-\frac{\xi}{\sqrt{\xi^2
		+\Delta_{\xi}^2}}\right),
	\end{equation}
where $N_0(E)$ is the normal-state DOS given by Eq.~(\ref{eq:DOS_d}) and $\Delta_{\xi}=\theta(\cutoff-|\xi|)\Delta$ with $\Delta$ the zero-temperature gap. We set $E_{0\alpha}=0$ and $m=m_{\alpha}$ as in Sec.~\ref{sec:1band} and move to dimensionless variables. In 2D we have at $T=0$,
	\begin{equation}\label{eq:zeroTn}
		\tilde{n}=\begin{cases}
			0 & \tilde{\mu}<-1 \\
			\frac{1}{2}\left(\tilde{\mu}+1+\sqrt{\tilde{\mu}^2+\tilde{\Delta}^2}-
			\sqrt{1+\tilde{\Delta}^2}\right)
			& -1<\tilde{\mu}<1 \\
			\tilde{\mu} & \tilde{\mu}>1.
		\end{cases}
	\end{equation}
The gap equation is obtained by replacing $\tanh(\cdots)$ by unity in Eq.~(\ref{eq:gap-equationa}). For one band in 2D we find
	\begin{equation}\label{eq:ZeroTgap}
		1/\coupling=\begin{cases}
			0 & \tilde{\mu}<-1 \\
			\frac{1}{2}\ln\left(\frac{1+\sqrt{1+\tilde{\Delta}^2}}
			{\sqrt{\tilde{\mu}^2+\tilde{\Delta}^2}-\tilde{\mu}}\right) & -1<\tilde{\mu}<1 \\
			\ln\left(\frac{1+\sqrt{1+\tilde{\Delta}^2}}{\tilde{\Delta}}\right)
			& \tilde{\mu}>1.
		\end{cases}
	\end{equation}
For $\tilde{\mu}>1$, the gap is independent of density and given by $\tilde{\Delta}=1/\sinh(1/\coupling)$; combined with the weak-coupling result (\ref{eq:tc2D-highn}), this yields the usual weak-coupling BCS ratio $\Delta/(k_{\text{B}}T_c)=\pi/e^{\gamma}$. In the low-density regime $\tilde{\mu}<1$, we eliminate $\tilde{\Delta}$ among Eqs.~(\ref{eq:zeroTn}) and (\ref{eq:ZeroTgap}) to find $\tilde{\mu}=\tilde{n}+(\tilde{n}-1)e^{-2/\coupling}$. Solving for the gap, we then arrive at:
	\begin{equation}\label{eq:Deltalown}
		\tilde{\Delta}=\frac{\sqrt{\tilde{n}+\tilde{n}(\tilde{n}-1)e^{-2/\coupling}}}
		{\sinh(1/\coupling)}\qquad(\tilde{n}<1).
	\end{equation}
Comparing Eqs.~(\ref{eq:Deltalown}) and (\ref{eq:tc2D-lown}) we see that $T_c$ increases faster than $\Delta$ with increasing density. As a result the gap to $T_c$ ratio vanishes for $n\to0$, as we show in Fig.~\ref{fig:ratio-2D}. This result may look surprising in view of the fact that known two-dimensional superconductors tend to have a gap to $T_c$ ratio larger than the BCS value. The suppression shown in Fig.~\ref{fig:ratio-2D} concerns a regime of density which none of these known superconductors has reached until now, to our knowledge.


\begin{thebibliography}{31}%
\makeatletter
\providecommand \@ifxundefined [1]{%
 \@ifx{#1\undefined}
}%
\providecommand \@ifnum [1]{%
 \ifnum #1\expandafter \@firstoftwo
 \else \expandafter \@secondoftwo
 \fi
}%
\providecommand \@ifx [1]{%
 \ifx #1\expandafter \@firstoftwo
 \else \expandafter \@secondoftwo
 \fi
}%
\providecommand \natexlab [1]{#1}%
\providecommand \enquote  [1]{``#1''}%
\providecommand \bibnamefont  [1]{#1}%
\providecommand \bibfnamefont [1]{#1}%
\providecommand \citenamefont [1]{#1}%
\providecommand \href@noop [0]{\@secondoftwo}%
\providecommand \href [0]{\begingroup \@sanitize@url \@href}%
\providecommand \@href[1]{\@@startlink{#1}\@@href}%
\providecommand \@@href[1]{\endgroup#1\@@endlink}%
\providecommand \@sanitize@url [0]{\catcode `\\12\catcode `\$12\catcode
  `\&12\catcode `\#12\catcode `\^12\catcode `\_12\catcode `\%12\relax}%
\providecommand \@@startlink[1]{}%
\providecommand \@@endlink[0]{}%
\providecommand \url  [0]{\begingroup\@sanitize@url \@url }%
\providecommand \@url [1]{\endgroup\@href {#1}{\urlprefix }}%
\providecommand \urlprefix  [0]{URL }%
\providecommand \Eprint [0]{\href }%
\providecommand \doibase [0]{http://dx.doi.org/}%
\providecommand \selectlanguage [0]{\@gobble}%
\providecommand \bibinfo  [0]{\@secondoftwo}%
\providecommand \bibfield  [0]{\@secondoftwo}%
\providecommand \translation [1]{[#1]}%
\providecommand \BibitemOpen [0]{}%
\providecommand \bibitemStop [0]{}%
\providecommand \bibitemNoStop [0]{.\EOS\space}%
\providecommand \EOS [0]{\spacefactor3000\relax}%
\providecommand \BibitemShut  [1]{\csname bibitem#1\endcsname}%
\let\auto@bib@innerbib\@empty
\bibitem [{\citenamefont {Bardeen}\ \emph {et~al.}(1957)\citenamefont
  {Bardeen}, \citenamefont {Cooper},\ and\ \citenamefont
  {Schrieffer}}]{Bardeen-1957}%
  \BibitemOpen
  \bibfield  {author} {\bibinfo {author} {\bibfnamefont {J.}~\bibnamefont
  {Bardeen}}, \bibinfo {author} {\bibfnamefont {L.~N.}\ \bibnamefont {Cooper}},
  \ and\ \bibinfo {author} {\bibfnamefont {J.~R.}\ \bibnamefont {Schrieffer}},\
  }\emph {\bibinfo {title} {Microscopic theory of superconductivity}},\ \href
  {\doibase 10.1103/PhysRev.108.1175} {\bibfield  {journal} {\bibinfo
  {journal} {Phys. Rev.}\ }\textbf {\bibinfo {volume} {108}},\ \bibinfo {pages}
  {1175} (\bibinfo {year} {1957})}\BibitemShut {NoStop}%
\bibitem [{\citenamefont {Eagles}(1967)}]{Eagles-1967}%
  \BibitemOpen
  \bibfield  {author} {\bibinfo {author} {\bibfnamefont {D.~M.}\ \bibnamefont
  {Eagles}},\ }\emph {\bibinfo {title} {Predicted Transition Temperatures of
  Very Thin Films and Whiskers of Superconducting Semiconductors---Application
  to SrTiO$_3$}},\ \href {\doibase 10.1103/PhysRev.164.489} {\bibfield
  {journal} {\bibinfo  {journal} {Phys. Rev.}\ }\textbf {\bibinfo {volume}
  {164}},\ \bibinfo {pages} {489} (\bibinfo {year} {1967})}\BibitemShut
  {NoStop}%
\bibitem [{\citenamefont {Thompson}\ and\ \citenamefont
  {Blatt}(1963)}]{Thompson-1963}%
  \BibitemOpen
  \bibfield  {author} {\bibinfo {author} {\bibfnamefont {C.~J.}\ \bibnamefont
  {Thompson}}\ and\ \bibinfo {author} {\bibfnamefont {J.~M.}\ \bibnamefont
  {Blatt}},\ }\emph {\bibinfo {title} {Shape resonances in superconductors - II
  simplified theory}},\ \href 
  {\doibase http://dx.doi.org/10.1016/S0375-9601(63)80003-1} {\bibfield  {journal}
  {\bibinfo  {journal} {Phys. Lett.}\ }\textbf {\bibinfo {volume} {5}},\
  \bibinfo {pages} {6} (\bibinfo {year} {1963})}\BibitemShut {NoStop}%
\bibitem [{\citenamefont {Romero-Berm{\'u}dez}\ and\ \citenamefont
  {Garc{\'\i}a-Garc{\'\i}a}(2014)}]{Romero-Bermudez-2014}%
  \BibitemOpen
  \bibfield  {author} {\bibinfo {author} {\bibfnamefont {A.}~\bibnamefont
  {Romero-Berm{\'u}dez}}\ and\ \bibinfo {author} {\bibfnamefont {A.~M.}\
  \bibnamefont {Garc{\'\i}a-Garc{\'\i}a}},\ }\emph {\bibinfo {title} {Shape
  resonances and shell effects in thin-film multiband superconductors}},\ \href
  {\doibase 10.1103/PhysRevB.89.024510} {\bibfield  {journal} {\bibinfo
  {journal} {Phys. Rev. B}\ }\textbf {\bibinfo {volume} {89}},\ \bibinfo
  {pages} {024510} (\bibinfo {year} {2014})}\BibitemShut {NoStop}%
\bibitem [{\citenamefont {Eagles}(1969{\natexlab{a}})}]{Eagles-1969a}%
  \BibitemOpen
  \bibfield  {author} {\bibinfo {author} {\bibfnamefont {D.~M.}\ \bibnamefont
  {Eagles}},\ }\emph {\bibinfo {title} {Effective Masses in Zr-Doped
  Superconducting Ceramic SrTiO$_3$}},\ \href 
  {\doibase 10.1103/PhysRev.178.668} {\bibfield  {journal} {\bibinfo  {journal} {Phys.
  Rev.}\ }\textbf {\bibinfo {volume} {178}},\ \bibinfo {pages} {668} (\bibinfo
  {year} {1969}{\natexlab{a}})}\BibitemShut {NoStop}%
\bibitem [{\citenamefont {Eagles}(1969{\natexlab{b}})}]{Eagles-1969b}%
  \BibitemOpen
  \bibfield  {author} {\bibinfo {author} {\bibfnamefont {D.~M.}\ \bibnamefont
  {Eagles}},\ }\emph {\bibinfo {title} {Possible Pairing without
  Superconductivity at Low Carrier Concentrations in Bulk and Thin-Film
  Superconducting Semiconductors}},\ \href {\doibase 10.1103/PhysRev.186.456}
  {\bibfield  {journal} {\bibinfo  {journal} {Phys. Rev.}\ }\textbf {\bibinfo
  {volume} {186}},\ \bibinfo {pages} {456} (\bibinfo {year}
  {1969}{\natexlab{b}})}\BibitemShut {NoStop}%
\bibitem [{\citenamefont {Lin}\ \emph {et~al.}(2014)\citenamefont {Lin},
  \citenamefont {Bridoux}, \citenamefont {Gourgout}, \citenamefont {Seyfarth},
  \citenamefont {Kr{\"a}mer}, \citenamefont {Nardone}, \citenamefont
  {Fauqu{\'e}},\ and\ \citenamefont {Behnia}}]{Lin-2014}%
  \BibitemOpen
  \bibfield  {author} {\bibinfo {author} {\bibfnamefont {X.}~\bibnamefont
  {Lin}}, \bibinfo {author} {\bibfnamefont {G.}~\bibnamefont {Bridoux}},
  \bibinfo {author} {\bibfnamefont {A.}~\bibnamefont {Gourgout}}, \bibinfo
  {author} {\bibfnamefont {G.}~\bibnamefont {Seyfarth}}, \bibinfo {author}
  {\bibfnamefont {S.}~\bibnamefont {Kr{\"a}mer}}, \bibinfo {author}
  {\bibfnamefont {M.}~\bibnamefont {Nardone}}, \bibinfo {author} {\bibfnamefont
  {B.}~\bibnamefont {Fauqu{\'e}}}, \ and\ \bibinfo {author} {\bibfnamefont
  {K.}~\bibnamefont {Behnia}},\ }\emph {\bibinfo {title} {Critical Doping for
  the Onset of a Two-Band Superconducting Ground State in
  SrTiO$_{3-\delta}$}},\ \href {\doibase 10.1103/PhysRevLett.112.207002}
  {\bibfield  {journal} {\bibinfo  {journal} {Phys. Rev. Lett.}\ }\textbf
  {\bibinfo {volume} {112}},\ \bibinfo {pages} {207002} (\bibinfo {year}
  {2014})}\BibitemShut {NoStop}%
\bibitem [{\citenamefont {Ahrens}\ \emph {et~al.}(2007)\citenamefont {Ahrens},
  \citenamefont {Merkle}, \citenamefont {Rahmati},\ and\ \citenamefont
  {Maier}}]{Ahrens-2007}%
  \BibitemOpen
  \bibfield  {author} {\bibinfo {author} {\bibfnamefont {M.}~\bibnamefont
  {Ahrens}}, \bibinfo {author} {\bibfnamefont {R.}~\bibnamefont {Merkle}},
  \bibinfo {author} {\bibfnamefont {B.}~\bibnamefont {Rahmati}}, \ and\
  \bibinfo {author} {\bibfnamefont {J.}~\bibnamefont {Maier}},\ }\emph
  {\bibinfo {title} {Effective masses of electrons in $n$-type SrTiO$_3$
  determined from low-temperature specific heat capacities}},\ \href 
  {\doibase 10.1016/j.physb.2007.01.008} {\bibfield  {journal} {\bibinfo  {journal}
  {Physica B: Cond. Mat.}\ }\textbf {\bibinfo {volume} {393}},\ \bibinfo
  {pages} {239} (\bibinfo {year} {2007})}\BibitemShut {NoStop}%
\bibitem [{\citenamefont {Perali}\ \emph {et~al.}(1996)\citenamefont {Perali},
  \citenamefont {Bianconi}, \citenamefont {Lanzara},\ and\ \citenamefont
  {Saini}}]{Perali-1996}%
  \BibitemOpen
  \bibfield  {author} {\bibinfo {author} {\bibfnamefont {A.}~\bibnamefont
  {Perali}}, \bibinfo {author} {\bibfnamefont {A.}~\bibnamefont {Bianconi}},
  \bibinfo {author} {\bibfnamefont {A.}~\bibnamefont {Lanzara}}, \ and\
  \bibinfo {author} {\bibfnamefont {N.~L.}\ \bibnamefont {Saini}},\ }\emph
  {\bibinfo {title} {The gap amplification at a shape resonance in a
  superlattice of quantum stripes: A mechanism for high $T_c$}},\ \href
  {\doibase 10.1016/0038-1098(96)00373-0} {\bibfield  {journal} {\bibinfo
  {journal} {Solid State Comm.}\ }\textbf {\bibinfo {volume} {100}},\ \bibinfo
  {pages} {181} (\bibinfo {year} {1996})}\BibitemShut {NoStop}%
\bibitem [{\citenamefont {Innocenti}\ \emph {et~al.}(2010)\citenamefont
  {Innocenti}, \citenamefont {Poccia}, \citenamefont {Ricci}, \citenamefont
  {Valletta}, \citenamefont {Caprara}, \citenamefont {Perali},\ and\
  \citenamefont {Bianconi}}]{Innocenti-2010}%
  \BibitemOpen
  \bibfield  {author} {\bibinfo {author} {\bibfnamefont {D.}~\bibnamefont
  {Innocenti}}, \bibinfo {author} {\bibfnamefont {N.}~\bibnamefont {Poccia}},
  \bibinfo {author} {\bibfnamefont {A.}~\bibnamefont {Ricci}}, \bibinfo
  {author} {\bibfnamefont {A.}~\bibnamefont {Valletta}}, \bibinfo {author}
  {\bibfnamefont {S.}~\bibnamefont {Caprara}}, \bibinfo {author} {\bibfnamefont
  {A.}~\bibnamefont {Perali}}, \ and\ \bibinfo {author} {\bibfnamefont
  {A.}~\bibnamefont {Bianconi}},\ }\emph {\bibinfo {title} {Resonant and
  crossover phenomena in a multiband superconductor: Tuning the chemical
  potential near a band edge}},\ \href {\doibase 10.1103/PhysRevB.82.184528}
  {\bibfield  {journal} {\bibinfo  {journal} {Phys. Rev. B}\ }\textbf {\bibinfo
  {volume} {82}},\ \bibinfo {pages} {184528} (\bibinfo {year}
  {2010})}\BibitemShut {NoStop}%
\bibitem [{\citenamefont {Innocenti}\ \emph {et~al.}(2011)\citenamefont
  {Innocenti}, \citenamefont {Caprara}, \citenamefont {Poccia}, \citenamefont
  {Ricci}, \citenamefont {Valletta},\ and\ \citenamefont
  {Bianconi}}]{Innocenti-2011}%
  \BibitemOpen
  \bibfield  {author} {\bibinfo {author} {\bibfnamefont {D.}~\bibnamefont
  {Innocenti}}, \bibinfo {author} {\bibfnamefont {S.}~\bibnamefont {Caprara}},
  \bibinfo {author} {\bibfnamefont {N.}~\bibnamefont {Poccia}}, \bibinfo
  {author} {\bibfnamefont {A.}~\bibnamefont {Ricci}}, \bibinfo {author}
  {\bibfnamefont {A.}~\bibnamefont {Valletta}}, \ and\ \bibinfo {author}
  {\bibfnamefont {A.}~\bibnamefont {Bianconi}},\ }\emph {\bibinfo {title}
  {Shape resonance for the anisotropic superconducting gaps near a Lifshitz
  transition: the effect of electron hopping between layers}},\ \href 
  {\doibase 10.1088/0953-2048/24/1/015012} {\bibfield  {journal} {\bibinfo  {journal}
  {Supercond. Sci. Technol.}\ }\textbf {\bibinfo {volume} {24}},\ \bibinfo
  {pages} {015012} (\bibinfo {year} {2011})}\BibitemShut {NoStop}%
\bibitem [{\citenamefont {Guidini}\ and\ \citenamefont
  {Perali}(2014)}]{Guidini-2014}%
  \BibitemOpen
  \bibfield  {author} {\bibinfo {author} {\bibfnamefont {A.}~\bibnamefont
  {Guidini}}\ and\ \bibinfo {author} {\bibfnamefont {A.}~\bibnamefont
  {Perali}},\ }\emph {\bibinfo {title} {Band-edge {BCS-BEC} crossover in a
  two-band superconductor: physical properties and detection parameters}},\
  \href {\doibase 10.1088/0953-2048/27/12/124002} {\bibfield  {journal}
  {\bibinfo  {journal} {Supercond. Sci. Technol.}\ }\textbf {\bibinfo {volume}
  {27}},\ \bibinfo {pages} {124002} (\bibinfo {year} {2014})}\BibitemShut
  {NoStop}%
\bibitem [{\citenamefont {Zubko}\ \emph {et~al.}(2011)\citenamefont {Zubko},
  \citenamefont {Gariglio}, \citenamefont {Gabay}, \citenamefont {Ghosez},\
  and\ \citenamefont {Triscone}}]{Zubko-2011}%
  \BibitemOpen
  \bibfield  {author} {\bibinfo {author} {\bibfnamefont {P.}~\bibnamefont
  {Zubko}}, \bibinfo {author} {\bibfnamefont {S.}~\bibnamefont {Gariglio}},
  \bibinfo {author} {\bibfnamefont {M.}~\bibnamefont {Gabay}}, \bibinfo
  {author} {\bibfnamefont {P.}~\bibnamefont {Ghosez}}, \ and\ \bibinfo {author}
  {\bibfnamefont {J.-M.}\ \bibnamefont {Triscone}},\ }\emph {\bibinfo {title}
  {Interface Physics in Complex Oxide Heterostructures}},\ \href 
  {\doibase 10.1146/annurev-conmatphys-062910-140445} {\bibfield  {journal} {\bibinfo
  {journal} {Ann. Rev. Cond. Mat. Phys.}\ }\textbf {\bibinfo {volume} {2}},\
  \bibinfo {pages} {141} (\bibinfo {year} {2011})}\BibitemShut {NoStop}%
\bibitem [{\citenamefont {Pentcheva}\ and\ \citenamefont
  {Pickett}(2010)}]{Pentcheva-2010}%
  \BibitemOpen
  \bibfield  {author} {\bibinfo {author} {\bibfnamefont {R.}~\bibnamefont
  {Pentcheva}}\ and\ \bibinfo {author} {\bibfnamefont {W.~E.}\ \bibnamefont
  {Pickett}},\ }\emph {\bibinfo {title} {Electronic phenomena at complex oxide
  interfaces: insights from first principles}},\ \href 
  {\doibase 10.1088/0953-8984/22/4/043001} {\bibfield  {journal} {\bibinfo  {journal} {J.
  Phys. Condens. Matter}\ }\textbf {\bibinfo {volume} {22}},\ \bibinfo {pages}
  {043001} (\bibinfo {year} {2010})}\BibitemShut {NoStop}%
\bibitem [{\citenamefont {Cancellieri}\ \emph {et~al.}(2014)\citenamefont
  {Cancellieri}, \citenamefont {Reinle-Schmitt}, \citenamefont {Kobayashi},
  \citenamefont {Strocov}, \citenamefont {Willmott}, \citenamefont {Fontaine},
  \citenamefont {Ghosez}, \citenamefont {Filippetti}, \citenamefont {Delugas},\
  and\ \citenamefont {Fiorentini}}]{Cancellieri-2014}%
  \BibitemOpen
  \bibfield  {author} {\bibinfo {author} {\bibfnamefont {C.}~\bibnamefont
  {Cancellieri}}, \bibinfo {author} {\bibfnamefont {M.~L.}\ \bibnamefont
  {Reinle-Schmitt}}, \bibinfo {author} {\bibfnamefont {M.}~\bibnamefont
  {Kobayashi}}, \bibinfo {author} {\bibfnamefont {V.~N.}\ \bibnamefont
  {Strocov}}, \bibinfo {author} {\bibfnamefont {P.~R.}\ \bibnamefont
  {Willmott}}, \bibinfo {author} {\bibfnamefont {D.}~\bibnamefont {Fontaine}},
  \bibinfo {author} {\bibfnamefont {P.}~\bibnamefont {Ghosez}}, \bibinfo
  {author} {\bibfnamefont {A.}~\bibnamefont {Filippetti}}, \bibinfo {author}
  {\bibfnamefont {P.}~\bibnamefont {Delugas}}, \ and\ \bibinfo {author}
  {\bibfnamefont {V.}~\bibnamefont {Fiorentini}},\ }\emph {\bibinfo {title}
  {Doping-dependent band structure of {LaAlO$_3$/SrTiO$_3$} interfaces by soft
  x-ray polarization-controlled resonant angle-resolved photoemission}},\ \href
  {\doibase 10.1103/PhysRevB.89.121412} {\bibfield  {journal} {\bibinfo
  {journal} {Phys. Rev. B}\ }\textbf {\bibinfo {volume} {89}},\ \bibinfo
  {pages} {121412} (\bibinfo {year} {2014})}\BibitemShut {NoStop}%
\bibitem [{\citenamefont {Berezinski{\v \i}}(1970)}]{Berezinskii-1970}%
  \BibitemOpen
  \bibfield  {author} {\bibinfo {author} {\bibfnamefont {V.~L.}\ \bibnamefont
  {Berezinski{\v \i}}},\ }\emph {\bibinfo {title} {Destruction of Long-range
  Order in One-dimensional and Two-dimensional Systems having a Continuous
  Symmetry Group I. Classical Systems}},\ \href
  {http://www.jetp.ac.ru/cgi-bin/e/index/e/32/3/p493?a=list} {\bibfield
  {journal} {\bibinfo  {journal} {Zh. Eksp. Teor. Fiz.}\ }\textbf {\bibinfo
  {volume} {59}},\ \bibinfo {pages} {907} (\bibinfo {year} {1970})}\BibitemShut
  {NoStop}%
\bibitem [{\citenamefont {Kosterlitz}\ and\ \citenamefont
  {Thouless}(1973)}]{Kosterlitz-1973}%
  \BibitemOpen
  \bibfield  {author} {\bibinfo {author} {\bibfnamefont {J.~M.}\ \bibnamefont
  {Kosterlitz}}\ and\ \bibinfo {author} {\bibfnamefont {D.~J.}\ \bibnamefont
  {Thouless}},\ }\emph {\bibinfo {title} {Ordering, metastability and phase
  transitions in two-dimensional systems}},\ \href
  {http://stacks.iop.org/0022-3719/6/i=7/a=010} {\bibfield  {journal} {\bibinfo
   {journal} {J. Phys. C: Solid State Phys.}\ }\textbf {\bibinfo {volume}
  {6}},\ \bibinfo {pages} {1181} (\bibinfo {year} {1973})}\BibitemShut
  {NoStop}%
\bibitem [{\citenamefont {Larkin}\ and\ \citenamefont
  {Varlamov}(2005)}]{Larkin-2005}%
  \BibitemOpen
  \bibfield  {author} {\bibinfo {author} {\bibfnamefont {A.}~\bibnamefont
  {Larkin}}\ and\ \bibinfo {author} {\bibfnamefont {A.}~\bibnamefont
  {Varlamov}},\ }\href {\doibase 10.1093/acprof:oso/9780198528159.001.0001}
  {\emph {\bibinfo {title} {Theory of Fluctuations in Superconductors}}},\
  \bibinfo {series} {International Series of Monographs on Physics}, Vol.\
  \bibinfo {volume} {127}\ (\bibinfo  {publisher} {Oxford University Press},\
  \bibinfo {year} {2005})\BibitemShut {NoStop}%
\bibitem [{\citenamefont {Leggett}(1980)}]{Leggett-1980}%
  \BibitemOpen
  \bibfield  {author} {\bibinfo {author} {\bibfnamefont {A.~J.}\ \bibnamefont
  {Leggett}},\ }\enquote {\bibinfo {title} {Diatomic molecules and {Cooper}
  pairs},}\ in\ \href {\doibase 10.1007/BFb0120125} {\emph {\bibinfo
  {booktitle} {Modern Trends in the Theory of Condensed Matter}}},\ \bibinfo
  {series} {Lecture Notes in Physics}, Vol.\ \bibinfo {volume} {115},\ \bibinfo
  {editor} {edited by\ \bibinfo {editor} {\bibfnamefont {A.}~\bibnamefont
  {P{\c{e}}kalski}}\ and\ \bibinfo {editor} {\bibfnamefont {J.~A.}\
  \bibnamefont {Przystawa}}}\ (\bibinfo  {publisher} {Springer-Verlag},\
  \bibinfo {address} {Berlin},\ \bibinfo {year} {1980})\ p.~\bibinfo {pages}
  {13}\BibitemShut {NoStop}%
\bibitem [{\citenamefont {Randeria}\ and\ \citenamefont
  {Taylor}(2014)}]{Randeria-2014}%
  \BibitemOpen
  \bibfield  {author} {\bibinfo {author} {\bibfnamefont {M.}~\bibnamefont
  {Randeria}}\ and\ \bibinfo {author} {\bibfnamefont {E.}~\bibnamefont
  {Taylor}},\ }\emph {\bibinfo {title} {Crossover from
  Bardeen-Cooper-Schrieffer to Bose-Einstein Condensation and the Unitary Fermi
  Gas}},\ \href {\doibase 10.1146/annurev-conmatphys-031113-133829} {\bibfield
  {journal} {\bibinfo  {journal} {Ann. Rev. Cond. Mat. Phys.}\ }\textbf
  {\bibinfo {volume} {5}},\ \bibinfo {pages} {209} (\bibinfo {year}
  {2014})}\BibitemShut {NoStop}%
\bibitem [{\citenamefont {Hainzl}\ and\ \citenamefont
  {Seiringer}(2008)}]{Hainzl-2008}%
  \BibitemOpen
  \bibfield  {author} {\bibinfo {author} {\bibfnamefont {C.}~\bibnamefont
  {Hainzl}}\ and\ \bibinfo {author} {\bibfnamefont {R.}~\bibnamefont
  {Seiringer}},\ }\emph {\bibinfo {title} {Critical temperature and energy gap
  for the BCS equation}},\ \href {\doibase 10.1103/PhysRevB.77.184517}
  {\bibfield  {journal} {\bibinfo  {journal} {Phys. Rev. B}\ }\textbf {\bibinfo
  {volume} {77}},\ \bibinfo {pages} {184517} (\bibinfo {year}
  {2008})}\BibitemShut {NoStop}%
\bibitem [{\citenamefont {van~der Marel}(1990)}]{vanderMarel-1990}%
  \BibitemOpen
  \bibfield  {author} {\bibinfo {author} {\bibfnamefont {D.}~\bibnamefont
  {van~der Marel}},\ }\emph {\bibinfo {title} {Anomalous behaviour of the
  chemical potential in superconductors with a low density of charge
  carriers}},\ \href {\doibase 10.1016/0921-4534(90)90429-I} {\bibfield
  {journal} {\bibinfo  {journal} {Physica C}\ }\textbf {\bibinfo {volume}
  {165}},\ \bibinfo {pages} {35} (\bibinfo {year} {1990})}\BibitemShut
  {NoStop}%
\bibitem [{Note1()}]{Note1}%
  \BibitemOpen
  \bibinfo {note} {We pull out a minus sign from the interaction for
  convenience. $V_{\alpha \beta }$ can be positive (attractive interaction) or
  negative (repulsive interaction).}\BibitemShut {Stop}%
\bibitem [{Note2()}]{Note2}%
  \BibitemOpen
  \bibinfo {note} {The standard numerical integration packages fail to produce
  an accurate result for Eq.~(\ref {eq:psi}), or take a prohibitively long time
  to converge, especially in the limit $b\to 0$ of interest to us. For a fast
  and accurate numerical evaluation of Eq.~(\ref {eq:psi}) we have used the
  representation of the Fermi function proposed in Ref.~\protect
  \rev@citealpnum {Osaki-2007}. With 14 Osaki poles and residues, we build a
  representation of the function $\protect \qopname \relax o{tanh}(x/2)$ which
  has an accuracy of $\sim 10^{-16}$ for all $x$, setting the function to $\pm
  1$ for $|x|>16\protect \qopname \relax o{ln}(10)$. With this representation,
  Eq.~(\ref {eq:psi}) can be evaluated analytically. This provides a very fast
  implementation of the functions $\psi _d(a,b)$ with double precision
  accuracy.}\BibitemShut {Stop}%
\bibitem [{\citenamefont {Ozaki}(2007)}]{Osaki-2007}%
  \BibitemOpen
  \bibfield  {author} {\bibinfo {author} {\bibfnamefont {T.}~\bibnamefont
  {Ozaki}},\ }\emph {\bibinfo {title} {Continued fraction representation of the
  {Fermi-Dirac} function for large-scale electronic structure calculations}},\
  \href {\doibase 10.1103/PhysRevB.75.035123} {\bibfield  {journal} {\bibinfo
  {journal} {Phys. Rev. B}\ }\textbf {\bibinfo {volume} {75}},\ \bibinfo
  {pages} {035123} (\bibinfo {year} {2007})}\BibitemShut {NoStop}%
\bibitem [{\citenamefont {Chubukov}\ \emph {et~al.}(2016)\citenamefont
  {Chubukov}, \citenamefont {Eremin},\ and\ \citenamefont
  {Efremov}}]{Chubukov-2016}%
  \BibitemOpen
  \bibfield  {author} {\bibinfo {author} {\bibfnamefont {A.~V.}\ \bibnamefont
  {Chubukov}}, \bibinfo {author} {\bibfnamefont {I.}~\bibnamefont {Eremin}}, \
  and\ \bibinfo {author} {\bibfnamefont {D.~V.}\ \bibnamefont {Efremov}},\
  }\emph {\bibinfo {title} {Superconductivity versus bound-state formation in a
  two-band superconductor with small {Fermi} energy: Applications to {Fe}
  pnictides/chalcogenides and doped {SrTiO$_3$}}},\ \href 
  {\doibase 10.1103/PhysRevB.93.174516} {\bibfield  {journal} {\bibinfo  {journal} {Phys.
  Rev. B}\ }\textbf {\bibinfo {volume} {93}},\ \bibinfo {pages} {174516}
  (\bibinfo {year} {2016})}\BibitemShut {NoStop}%
\bibitem [{\citenamefont {Zehetmayer}(2013)}]{Zehetmayer-2013}%
  \BibitemOpen
  \bibfield  {author} {\bibinfo {author} {\bibfnamefont {M.}~\bibnamefont
  {Zehetmayer}},\ }\emph {\bibinfo {title} {A review of two-band
  superconductivity: materials and effects on the thermodynamic and reversible
  mixed-state properties}},\ \href {\doibase 10.1088/0953-2048/26/4/043001}
  {\bibfield  {journal} {\bibinfo  {journal} {Supercond. Sci. Tech.}\ }\textbf
  {\bibinfo {volume} {26}},\ \bibinfo {pages} {043001} (\bibinfo {year}
  {2013})}\BibitemShut {NoStop}%
\bibitem [{\citenamefont {Suhl}\ \emph {et~al.}(1959)\citenamefont {Suhl},
  \citenamefont {Matthias},\ and\ \citenamefont {Walker}}]{Suhl-1959}%
  \BibitemOpen
  \bibfield  {author} {\bibinfo {author} {\bibfnamefont {H.}~\bibnamefont
  {Suhl}}, \bibinfo {author} {\bibfnamefont {B.~T.}\ \bibnamefont {Matthias}},
  \ and\ \bibinfo {author} {\bibfnamefont {L.~R.}\ \bibnamefont {Walker}},\
  }\emph {\bibinfo {title} {Bardeen-Cooper-Schrieffer Theory of
  Superconductivity in the Case of Overlapping Bands}},\ \href 
  {\doibase 10.1103/PhysRevLett.3.552} {\bibfield  {journal} {\bibinfo  {journal} {Phys.
  Rev. Lett.}\ }\textbf {\bibinfo {volume} {3}},\ \bibinfo {pages} {552}
  (\bibinfo {year} {1959})}\BibitemShut {NoStop}%
\bibitem [{\citenamefont {Bussmann-Holder}\ \emph {et~al.}(2004)\citenamefont
  {Bussmann-Holder}, \citenamefont {Micnas},\ and\ \citenamefont
  {Bishop}}]{Bussmann-Holder-2004}%
  \BibitemOpen
  \bibfield  {author} {\bibinfo {author} {\bibfnamefont {A.}~\bibnamefont
  {Bussmann-Holder}}, \bibinfo {author} {\bibfnamefont {R.}~\bibnamefont
  {Micnas}}, \ and\ \bibinfo {author} {\bibfnamefont {A.~R.}\ \bibnamefont
  {Bishop}},\ }\emph {\bibinfo {title} {Enhancements of the superconducting
  transition temperature within the two-band model}},\ \href 
  {\doibase 10.1140/epjb/e2004-00065-5} {\bibfield  {journal} {\bibinfo  {journal} {Eur.
  Phys. J. B}\ }\textbf {\bibinfo {volume} {37}},\ \bibinfo {pages} {345}
  (\bibinfo {year} {2004})}\BibitemShut {NoStop}%
\bibitem [{\citenamefont {Fernandes}\ \emph {et~al.}(2013)\citenamefont
  {Fernandes}, \citenamefont {Haraldsen}, \citenamefont {W{\"o}lfle},\ and\
  \citenamefont {Balatsky}}]{Fernandes-2013}%
  \BibitemOpen
  \bibfield  {author} {\bibinfo {author} {\bibfnamefont {R.~M.}\ \bibnamefont
  {Fernandes}}, \bibinfo {author} {\bibfnamefont {J.~T.}\ \bibnamefont
  {Haraldsen}}, \bibinfo {author} {\bibfnamefont {P.}~\bibnamefont
  {W{\"o}lfle}}, \ and\ \bibinfo {author} {\bibfnamefont {A.~V.}\ \bibnamefont
  {Balatsky}},\ }\emph {\bibinfo {title} {Two-band superconductivity in doped
  SrTiO$_3$ films and interfaces}},\ \href 
  {\doibase 10.1103/PhysRevB.87.014510} {\bibfield  {journal} {\bibinfo  {journal} {Phys.
  Rev. B}\ }\textbf {\bibinfo {volume} {87}},\ \bibinfo {pages} {014510}
  (\bibinfo {year} {2013})}\BibitemShut {NoStop}%
\bibitem [{\citenamefont {Caviglia}\ \emph {et~al.}(2008)\citenamefont
  {Caviglia}, \citenamefont {Gariglio}, \citenamefont {Reyren}, \citenamefont
  {Jaccard}, \citenamefont {Schneider}, \citenamefont {Gabay}, \citenamefont
  {Thiel}, \citenamefont {Hammerl}, \citenamefont {Mannhart},\ and\
  \citenamefont {Triscone}}]{Caviglia-2008}%
  \BibitemOpen
  \bibfield  {author} {\bibinfo {author} {\bibfnamefont {A.~D.}\ \bibnamefont
  {Caviglia}}, \bibinfo {author} {\bibfnamefont {S.}~\bibnamefont {Gariglio}},
  \bibinfo {author} {\bibfnamefont {N.}~\bibnamefont {Reyren}}, \bibinfo
  {author} {\bibfnamefont {D.}~\bibnamefont {Jaccard}}, \bibinfo {author}
  {\bibfnamefont {T.}~\bibnamefont {Schneider}}, \bibinfo {author}
  {\bibfnamefont {M.}~\bibnamefont {Gabay}}, \bibinfo {author} {\bibfnamefont
  {S.}~\bibnamefont {Thiel}}, \bibinfo {author} {\bibfnamefont
  {G.}~\bibnamefont {Hammerl}}, \bibinfo {author} {\bibfnamefont
  {J.}~\bibnamefont {Mannhart}}, \ and\ \bibinfo {author} {\bibfnamefont
  {J.-M.}\ \bibnamefont {Triscone}},\ }\emph {\bibinfo {title} {Electric field
  control of the {$LaAlO_3$/$SrTiO_3$} interface ground state}},\ \href
  {\doibase 10.1038/nature07576} {\bibfield  {journal} {\bibinfo  {journal}
  {Nature}\ }\textbf {\bibinfo {volume} {456}},\ \bibinfo {pages} {624}
  (\bibinfo {year} {2008})}\BibitemShut {NoStop}%
\end{thebibliography}
\end{document}